\documentclass[aps,prresearch,twocolumn,showpacs,super-scriptaddress,groupedaddress,acknowledgment]{revtex4-1}  % for review and submission
\usepackage{dcolumn}   % needed for some tables
\usepackage{bm}        % for math

\usepackage{hyperref}
\hypersetup{
}

\usepackage{amsmath}
\usepackage{braket}
\usepackage{amsfonts}
\usepackage{graphicx}
\usepackage{amsthm}
\usepackage{color}
\usepackage{setspace}
\usepackage{amssymb}
\usepackage{latexsym}
\usepackage{here,comment}

\theoremstyle{definition}

\newtheorem*{theorem*}{Theorem}

\newtheorem*{definition*}{Definition}

\newcommand{\av}[1]{\overline{#1}}
\newcommand{\mc}[1]{\mathcal{#1}}
\newcommand{\mr}[1]{\mathrm{#1}}
\newcommand{\mbb}[1]{\mathbb{#1}}

\newcommand{\hH}{\hat{H}}

\newcommand{\hrho}{\hat{\rho}}
\newcommand{\lrs}[1]{\left( #1 \right)}
\newcommand{\lrm}[1]{{\left\{ #1 \right\}}}
\newcommand{\lrl}[1]{\left[ #1 \right]}
\newcommand{\lrv}[1]{\left| #1 \right|}

\newcommand{\fracd}[2]{\frac{\mathrm{d} #1 }{\mathrm{d} #2 }}

\newcommand{\aln}[1]{
\begin{align}
#1
\end{align}
}

\newcommand{\MATh}[4]{
\begin{pmatrix}
#1 & #2 \\
#3 & #4 \\
\end{pmatrix}
}

\newcommand{\ra}{\rightarrow}

\newcommand{\Tr}{\mr{Tr}}

\begin{document}
\title{
Universality classes of non-Hermitian random matrices
}
\date{\today}
\author{Ryusuke Hamazaki$^{1,2}$, Kohei Kawabata$^1$, Naoto Kura$^1$, and Masahito Ueda$^{1,3}$}
\affiliation{
$^1$Department of Physics, University of Tokyo, 7-3-1 Hongo, Bunkyo-ku, Tokyo 113-0033, Japan\\
$^2$Nonequilibrium Quantum Statistical Mechanics RIKEN Hakubi Research Team,
RIKEN Cluster for Pioneering Research (CPR), RIKEN iTHEMS, Wako, Saitama 351-0198, Japan\\
$^3$RIKEN Center for Emergent Matter Science (CEMS), Wako 351-0198, Japan
}

\begin{abstract}

Non-Hermitian random matrices have been utilized in such diverse  fields as dissipative and stochastic processes, mesoscopic physics, nuclear physics, and neural networks. However, the only known universal level-spacing statistics is that of the Ginibre ensemble characterized by complex-conjugation symmetry.
Here we report our discovery of two other distinct universality classes characterized by transposition symmetry.
We find that transposition symmetry alters repulsive interactions between two neighboring eigenvalues and deforms their spacing distribution.
Such alteration is not possible with other symmetries including Ginibre's complex-conjugation symmetry which can affect only nonlocal correlations. Our results complete the non-Hermitian counterpart of Wigner-Dyson's threefold universal statistics of Hermitian random matrices and serve as a basis for characterizing nonintegrability and chaos in open quantum systems with  symmetry.

\end{abstract}

\maketitle

\section{Introduction}
\begin{figure*}
\begin{center}
\includegraphics[width=\linewidth]{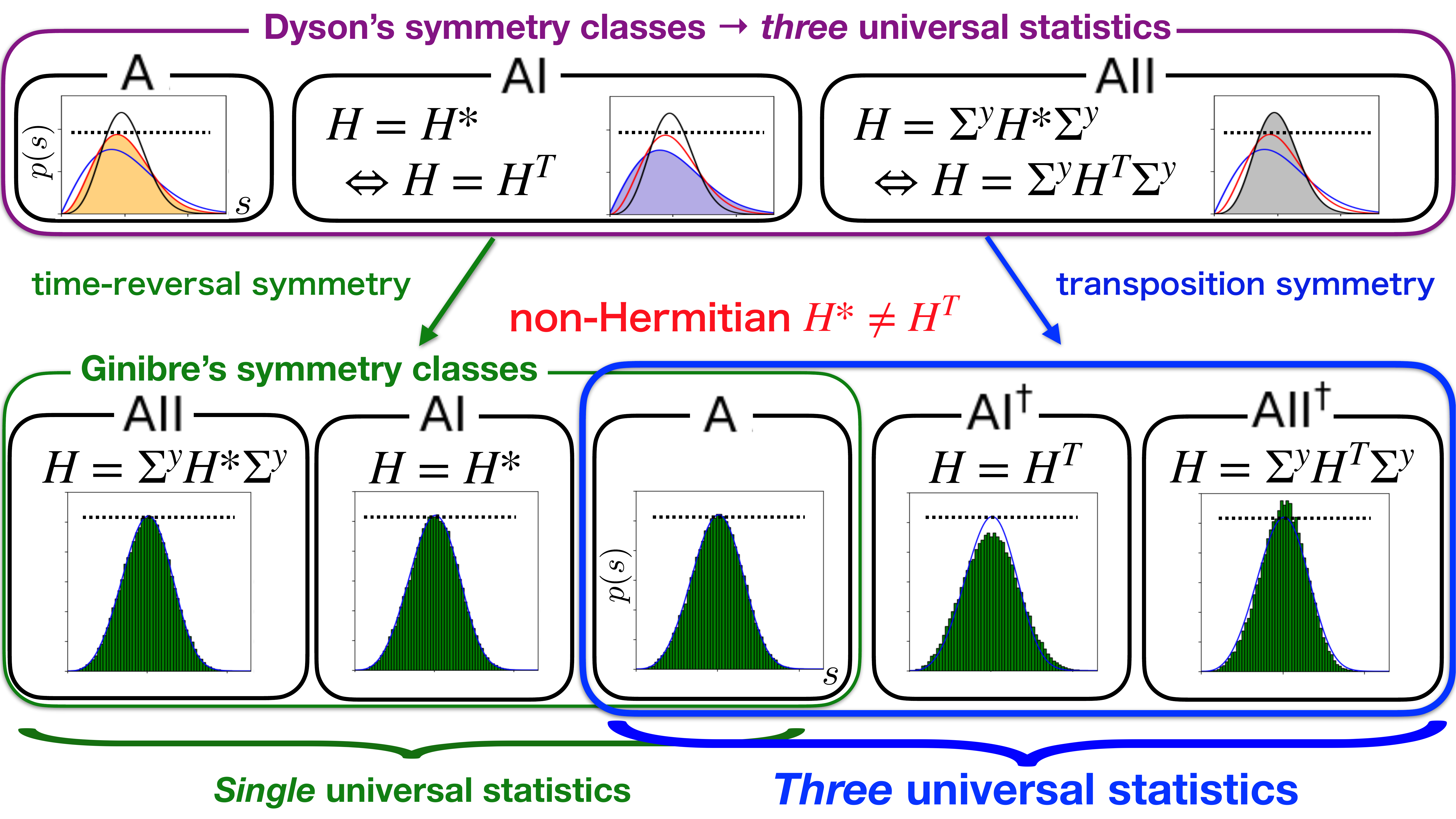}
\caption{
Schematic illustration of  universality classes of random matrices.
Dyson's threefold symmetry classes of Hermitian matrices in terms of time-reversal symmetry (TRS) lead to three distinct universal statistics of level-spacing distributions $p(s)$, where $\Sigma^y$ is defined in Eq. (3). In each panel, the level-spacing distributions corresponding to GSE, GUE and GOE are shown from top to bottom.
In contrast, Ginibre's three classes of non-Hermitian matrices based on TRS lead to  a single universality class.
We show that three symmetry classes in terms of transposition symmetry, which is distinct from TRS due to non-Hermiticity, lead to three distinct universality classes of $p(s)$, as seen from  the peak structure as  in the Hermitian case.
The black dotted lines show the peak height for class A in the Hermitian and non-Hermitian cases, and the blue solid curve in each panel in the non-Hermitian case shows Ginibre's level-spacing distribution.
}
\label{fig:fig1}
\end{center}
\end{figure*}

Symmetry and universality are two important concepts of Hermitian random matrix theory.
Dyson's threefold symmetry classes in terms of time-reversal symmetry (TRS)~\cite{Dyson62T} led to three distinct universality classes (Fig.~\ref{fig:fig1}),  where certain spectral statistics become independent of the detailed structures of matrices~\cite{Guhr98}.
The three distinct universal statistics have found applications in diverse research fields including nuclear physics~\cite{Brody81},  mesoscopic physics~\cite{Beenakker97,Beenakker15}, quantum chaos~\cite{Borgonovi16,DAlessio16}, and information theory~\cite{Tulino04}.
The most direct manifestation of the universality lies in the
level-spacing distribution, which measures local correlations of eigenvalues.
Level-spacing distributions of nonintegrable systems are described by those of Gaussian random matrices known as Wigner-Dyson's universal statistics~\cite{Bohigas84,DAlessio16,Assmann16}.
They belong to one of the following three random matrix ensembles: the Gaussian unitary ensemble (GUE) for the class without TRS (class A), the Gaussian orthogonal ensemble (GOE) for the class with TRS whose square is $+1$ (class AI), and the Gaussian symplectic ensemble (GSE) for the class with TRS whose square is $-1$ (class AII).

\begin{table}[b]
  \caption{
 Symmetry classes, constraints, and nearest-neighbor spacing distributions.
  The sign of each symmetry class indicates whether the square of the symmetry  operator is $+1$ or $-1$.
  Only classes AI$^\dag$ and AII$^\dag$ show the distributions different from that of Ginibre's unitary ensemble (GinUE). Here TRS, PHS, CS, SLS and pH stand for time-reversal symmetry, particle-hole symmetry, chiral symmetry, sublattice symmetry and pseudo-Hermiticity, respectively.}

\label{tabel}
\centering
  \begin{tabular}{cccc} \hline\hline
    Class &Symmetry& Constraint  & $p_\mr{GinUE}(s)$?\\ \hline
    A & None  &-&Yes~\cite{Ginibre65}  \\
    AI (D$^\dag$) &TRS, $+$ (PHS$^\dag$, $+$)&$H=H^*$  &Yes~\cite{Grobe89} \\
    AII (C$^\dag$)&TRS, $-$ (PHS$^\dag$, $-$) &$H=\Sigma^yH^*\Sigma^y$& Yes~\cite{Ginibre65}  \\
    AI$^\dag$ &TRS$^\dag$, $+$&$H=H^T$  & \textbf{No} \\
    AII$^\dag$ &TRS$^\dag$, $-$&$H=\Sigma^yH^T\Sigma^y$ & \textbf{No}  \\
		    D &PHS, $+$ &$H=-H^T$&Yes\\
    C &PHS, $-$ &$H=-\Sigma^yH^T\Sigma^y$& Yes\\
		AIII & CS (pH)&$H=-\Sigma^zH^\dag\Sigma^z$ &Yes \\
		AIII$^\dag$ & SLS (CS$^\dag$)&$H=-\Sigma^zH\Sigma^z$ &Yes  \\ \hline\hline

  \end{tabular}
	\label{table1}
\end{table}

While Dyson's classification is about Hermitian matrices,
non-Hermiticity plays a key role in such diverse systems  as dissipative systems~\cite{Grobe88,Grobe89,Chalker97}, mesoscopic systems~\cite{Schomerus17}, and neural networks~\cite{Sommers88}.
Many of these systems have been investigated in terms of non-Hermitian random-matrix ensembles introduced by Ginibre (Fig.~\ref{fig:fig1}), which are referred to as GinUE, GinOE, and GinSE as non-Hermitian extensions of GUE, GOE, and GSE~\cite{Ginibre65}.
These three Gaussian ensembles are defined in terms of complex conjugation (TRS) and have thoroughly been investigated~\cite{Lehmann91,Edelman95,Chalker98,Kanzieper05,Forrester07,Burda14}.

Interestingly, three  different symmetry classes (GinUE, GinOE, and GinSE for the Gaussian case) share the same universal level-spacing statistics of GinUE unlike the Hermitian case.
In fact, TRS only creates nonlocal correlations between complex-conjugate pairs of eigenvalues, but does not alter the repulsive interactions between neighboring eigenvalues away from the real axis.
Thus, the Ginibre distribution is the only previously known universal nearest-neighbor spacing distribution that is common to all three different symmetry classes with TRS, in contrast to the Hermitian case where TRS leads to three distinct universality classes.
Here, symmetry classification is defined solely by algebraic structures of matrix ensembles, whereas the universality classification is defined by spectral statistics that may be the same for different matrix ensembles~\cite{Guhr98}.

Motivated by experiments of non-Hermitian systems~\cite{Guo09,Ruter10,Regensburger12,Peng14,Assawaworrarit17,Fleury15,Bender13,Gao15,Xu16,Li19},  symmetry classification of non-Hermitian matrix ensembles has witnessed remarkable developments~\cite{Bernard02,Magnea08,Kawabata19S}.
Quite recently, two of us have shown that there are 38 non-Hermitian symmetry classes~\cite{Kawabata19S} as an extension of the Hermitian symmetry classes by Altland and Zirnbauer~\cite{Altland97}.
The classification includes symmetry classes  AI$^{\dag}$ and AII$^{\dag}$ that arise from transposition symmetry~\cite{Bernard02, Hastings01,Magnea08,Esaki11,Sato12T,Forrester16,Lieu18,Budich18,Kawabata19S,Zhou19}, which is Hermitian conjugate of TRS, i.e, TRS$^{\dag}$.
 A rich variety of experimentally realizable systems belongs to these classes such as optical systems with gain and/or loss~\cite{Konotop16,ElGanainy18,Miri19}.
However, it has remained largely unexplored how these symmetry classes are related to the universality classification of non-Hermitian random matrices.
While  some spectral statistics have been calculated for non-Ginibre Gaussian random matrices~\cite{Hastings01,Forrester16},   their universality has yet to be explored.
Note that, while Ref.~\cite{Hastings01} claims that it
exhausts the universality classification of non-{H}ermitian random matrices,
the result presented there is not universal statistics but symmetry classification
in terms of symmetric spaces.
We also note that, level-spacing distributions of real eigenvalues are sensitive to TRS~\cite{Graefe15}, and that  different universality classes can appear in the limit of weak non-{H}ermiticity~\cite{Fyodorov98, Sommers99}, where eigenvalues appear near the real axis. We do not consider these special  cases here.

We here investigate the nearest-neighbor spacing distributions of random matrices which belong to the simplest classes (i.e., classes with a single symmetry) in the non-Hermitian symmetry classification~\cite{Bernard02,Magnea08,Kawabata19S}.
As shown in Table~\ref{table1} and Fig.~\ref{fig:fig2}(c-k), we numerically find two universal level-spacing distributions that are distinct from GinUE for classes AI$^\dag$ and AII$^\dag$. Thus,
Wigner-Dyson's three universal statistics are extended to the non-Hermitian three universal statistics (classes A, AI$^\dag$, and  AII$^\dag$) with transposition symmetry (TRS$^\dag$) instead of complex-conjugation symmetry (TRS) in Ginibre's symmetry classes.
Analyzing  the level-spacing distributions for small matrices, we elucidate the similarity and difference between Dyson's classification and the new non-Hermitian generalization: TRS$^{\dag}$ alters the complex-valued degrees of freedom that control the repulsive interaction of neighboring eigenvalues and hence nearest-neighbor spacing distributions.
We show for the first time that the newly found universality classes indeed manifest themselves in dissipative quantum many-body nonintegrable systems described by the Lindblad dynamics and non-Hermitian Hamiltonians (see Sec.~\ref{sec5} for details).
%Our results serve as a basis for characterizing nonintegrability and chaos in open quantum systems.

The rest of this paper is organized as follows.
In Chapter~\ref{sec2}, we briefly review 9 non-Hermitian symmetry classes characterized by a single symmetry among the 38 symmetry classes~\cite{Kawabata19S}.
In Chapter~\ref{sec3}, we illustrate our finding of new universal level-spacing statistics arising from the transposition symmetry.
In Chapter~\ref{sec4}, we investigate effective small matrices that capture the qualitative features of the level-spacing distributions.
In Chapter~\ref{sec5}, we demonstrate that the newly found universality classes manifest themselves in dissipative quantum many-body systems described by the Lindblad dynamics and non-Hermitian Hamiltonians.
In Chapter~\ref{sec6}, we summarize the paper and discuss some future problems.

\section{Non-Hermitian ramification of symmetry classes}\label{sec2}
\subsection{Dyson and Ginibre's classifications}
Dyson's classification includes symmetry classes~A, AI, and AII.
Let us consider Hermitian $N\times N$ matrices in each of these three classes.
Matrices in class A have no symmetry constraint except  for non-Hermiticity.
Matrices in class AI satisfy
\aln{
H=H^*=H^T.
}
Matrices in class AII satisfy
 \aln{
 H=\Sigma^yH^*\Sigma^y=\Sigma^yH^T\Sigma^y,
 }
 where $H$ has a two-by-two block structure and
 \aln{
 \Sigma^y=\MATh{0}{-i\mbb{I}_{\frac{N}{2}\times \frac{N}{2}}}{i\mbb{I}_{\frac{N}{2}\times \frac{N}{2}}}{0}
 }
with the $N/2\times N/2$ identity matrix $\mbb{I}_{\frac{N}{2}\times \frac{N}{2}}$.
Here, without loss of generality, we take a unitary operator $\mc{T}$ as the identity in  $ H=\mc{T}H^*\mc{T}^{-1}$ with $\mc{T}\mc{T}^*=+1$ for class AI.
Similarly, $\mc{T}$ can be taken as $\Sigma^y$ in  $ H=\mc{T}H^*\mc{T}^{-1}$ with $\mc{T}\mc{T}^*=-1$ for class AII.
Below we consider the same simplification for the other classes (see Appendix~\ref{appA} for  general symmetry operators).

Ginibre presented the three non-Hermitian symmetry classes A, AI, and AII by considering complex-conjugation symmetry (TRS)~\cite{Ginibre65}.
Non-Hermitian random matrices in class A have no symmetry constraint; matrices in class AI respect
\aln{
H=H^*,
}
and matrices in class AII respect
 \aln{
 H=\Sigma^yH^*\Sigma^y.
 }
In the presence of TRS, eigenvalues are either real or form  complex-conjugate pairs
$
(E_{\alpha}, E^{*}_{\alpha}).
$

\subsection{The other symmetry classes with a single symmetry}
As shown in the next chapter, we find the non-Hermitian two universal statistics distinct from Ginibre's, by noticing that the distinction between complex conjugation and transposition for non-Hermitian matrices alters interactions between eigenvalues.
The three universality classes arise from symmetry classes AI$^\dag$ and AII$^\dag$ in addition to class A, where matrices respect transposition symmetry (TRS$^{\dag}$)~\cite{Kawabata19S}.
Matrices satisfy
\aln{
H=H^{T}\:(\neq H^*)
}
in class AI$^{\dag}$ and
\aln{
H=\Sigma^{y} H^{T} \Sigma^{y}\:(\neq \Sigma^{y} H^{*} \Sigma^{y})
}
 in class AII$^{\dag}$.
TRS$^\dag$ imposes constraints on the left and right eigenvectors of all the individual complex eigenvalues instead of nonlocal correlations.
In class AI$^{\dag}$, for example, a right eigenvector and the complex conjugate of the corresponding left eigenvector are proportional to each other.
On the other hand, TRS imposes not local constraints on individual eigenvectors but nonlocal constraints on complex-conjugate pairs of eigenvectors.

We can also investigate the entire non-Hermitian 38-fold classification~\cite{Kawabata19S}.
We focus on 9 symmetry classes that have single symmetries,  as summarized in Table~\ref{table1} (see Appendix~\ref{appA} for details).
In class D (resp. C), matrices possess particle-hole symmetry (PHS)  and satisfy $H^T=-H$ (resp. $\Sigma^yH^T\Sigma^y=-H$).
Class AIII has chiral symmetry (CS), which satisfies $\Sigma^z H^\dag\Sigma^z=-H$, where
\aln{
 \Sigma^z=\MATh{\mbb{I}_{\frac{N}{2}\times \frac{N}{2}}}{0}{0}{-\mbb{I}_{\frac{N}{2}\times \frac{N}{2}}}.
}
Finally, in class AIII$^\dag$ (Hermitian conjugate of class AIII),  sublattice symmetry (SLS)  exists and matrices satisfy $\Sigma^z H\Sigma^z=-H$.
Note that these symmetries  only create nonlocal correlations between pairs of complex eigenvalues.

\begin{figure*}[t]
\begin{center}
\includegraphics[width=\linewidth]{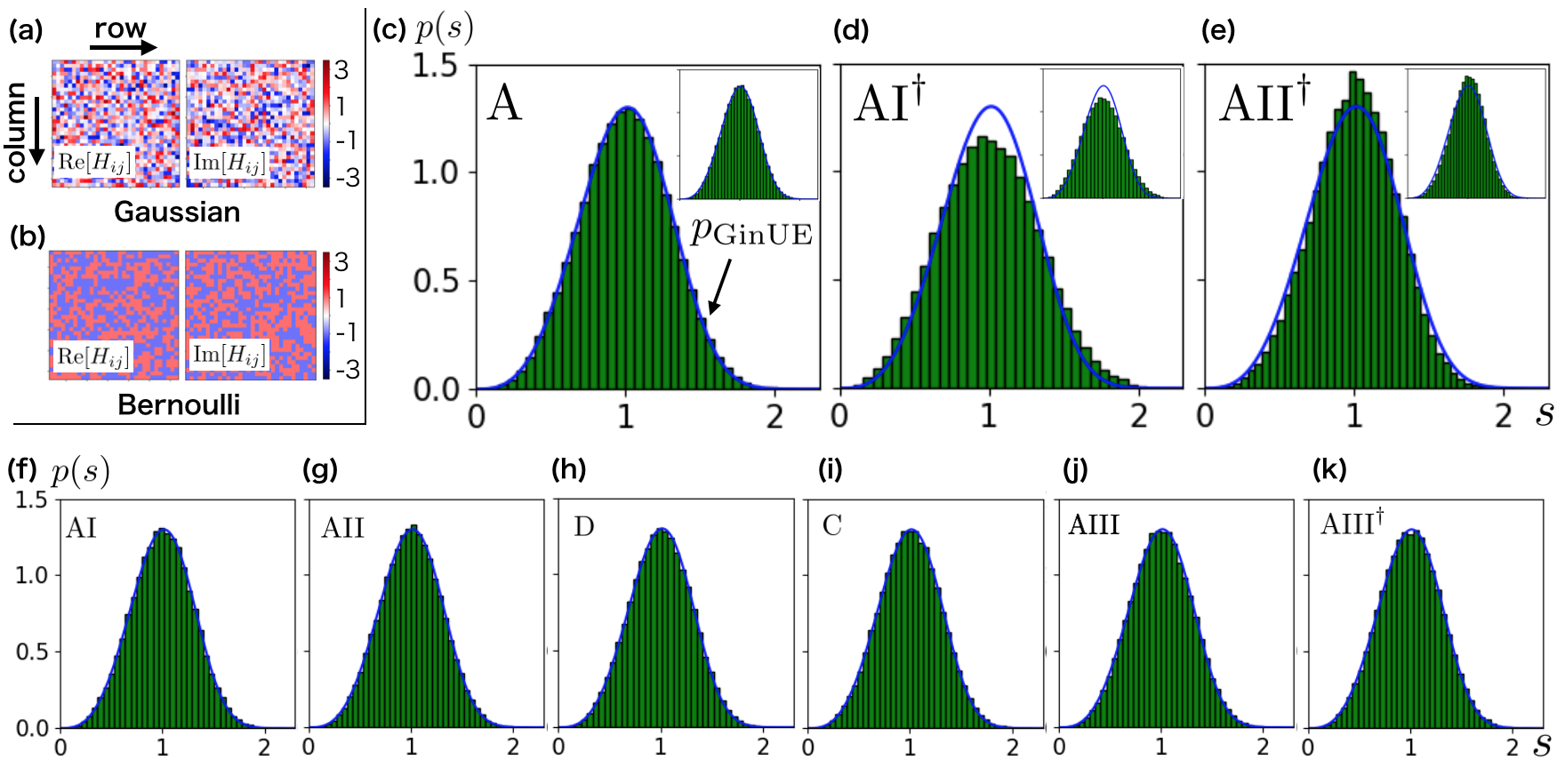}
\caption{
(a,b) Single realization of a matrix sampled from  (a) Gaussian and (b) Bernoulli random 32 $\times$ 32 matrices for class A, where the real and imaginary parts are displayed separately.
(c-k) Nearest-neighbor spacing distributions $p(s)$ of 2000 $\times$ 2000 random matrices for classes (c) A, (d) AI$^\dag$,  (e) AII$^\dag$, (f) AI, (g) AII, (h) D, (i) C, (j) AIII, and (k) AIII$^\dag$.
The main panels are for Gaussian matrices and the insets for classes A, AI$^\dag$, AII$^\dag$ are for Bernoulli matrices.
While classes A, AI, AII, D, C, AIII, and AIII$^\dag$ obey the GinUE distribution $p_\mr{GinUE}(s)$,
the peak of $p(s)$ is lower for class AI$^\dag$  and higher for class AII$^\dag$ than that of $p_\mr{GinUE}(s)$ shown in blue curves.
Statistics are taken from eigenvalues away from the edges of the spectrum (i.e., away from the circumference of a circle within which the spectrum resides) and away from the symmetric line (real or imaginary axis).
}
\label{fig:fig2}
\end{center}
\end{figure*}

\section{Level-spacing distributions}\label{sec3}
\subsection{New universality classes of the nearest-neighbor spacing distribution}
We now numerically calculate the nearest-neighbor spacing distributions $p(s)$ in the complex plane away from the symmetric line (i.e., the real or imaginary axis) for random matrices in each symmetry class,
where $p(s)$ is assumed to be normalized as
 \aln{
 \int_0^\infty p(s)ds=\int_0^\infty sp(s)ds =1.
 }
Note that the level spacing for an eigenvalue $E_\alpha$ on the complex plane is essentially given by the minimum distance $d_{1,\alpha}=\min_\beta|E_\alpha-E_\beta|$.
On the other hand, we need to perform the unfolding procedure to obtain the normalized level spacing $s_\alpha$ from the bare distance $d_{1,\alpha}$.
We do this by following Ref.~\cite{Haake}.
We first note that a local mean density of  eigenvalues can be evaluated as
\aln{
\av{\rho}=\frac{n}{\pi d_{n,\alpha}^2},
}
where $n$ is sufficiently larger than unity (say 10-30 in our simulations, which is sufficiently small compared with matrix sizes) and $d_{n,\alpha}$ is the $n$th nearest neighbor distance from $E_\alpha$.
Then, $s_\alpha$ is defined as~\cite{Haake}
\aln{
s_\alpha={d_{1,\alpha}}{\sqrt{\av{\rho}}},
}
with which the dependence of local density of eigenvalues vanishes.

To confirm the universality, we introduce Gaussian and Bernoulli ensembles of random matrices $H$.
For the Gaussian ensemble, the probability distribution takes the form
\aln{
P(H)dH \propto \exp(-\beta \mr{Tr}[H^\dag H])dH
}
with a constant $\beta>0$.
For the Bernoulli ensemble, each of matrix elements is  randomly taken from $\pm 1$ or $\pm 1\pm i$ while satisfying the symmetry constraint (see Appendix~\ref{appA}).
We show in Fig.~\ref{fig:fig2}(a,b)   single realizations of matrices sampled from the Gaussian and Bernoulli random matrices for  class A.

Figure~\ref{fig:fig2}(c-k) shows $p(s)$ for $2000\times 2000$ Gaussian or Bernoulli random matrices (the latter results are shown only for classes A, AI$^\dag$, and AII$^\dag$ for simplicity).
Class A follows the distributions of GinUE~\cite{Haake,Schomerus17},
\aln{
p_\mr{GinUE}(s)=C\tilde{p}(Cs),
}
where
\aln{
\tilde{p}(s)=\lim_{N\ra\infty}\lrl{\prod_{n=1}^{N-1}e_n(s^2)e^{-s^2}}  \sum_{n=1}^{N-1}\frac{2s^{2n+1}}{n!e_n(s^2)}
}
 with
 \aln{
 e_n(x)=\sum_{m=0}^n\frac{x^m}{m!}
 } and~\cite{Haake}
 \aln{
 C=\int_0^\infty dss\tilde{p}(s)=1.1429\cdots.
 }
We find similar results for classes AI, AII, D, C, AIII, and AIII$^\dag$, which indicates that, while symmetries for these classes create pairs of eigenvalues, i.e., $(E_\alpha, E_\alpha^*)$, $(E_\alpha,-E_\alpha)$, or $(E_\alpha,-E_\alpha^*)$, they do not alter  local correlations between neighboring eigenvalues away from the symmetric line (the real or imaginary axis).

By contrast, $p(s)$  for classes AI$^\dag$ and AII$^\dag$ are distinct from that of the GinUE in that the peak is lower (higher) and the variance is larger (smaller) for class AI$^{\dag}$ (AII$^{\dag}$) than that of the GinUE.
This is reminiscent of the Hermitian case, where the peak of the nearest-neighbor spacing distribution is lower for class AI (GOE) and higher for class AII (GSE) than class A (GUE).
Here, we note that the nearest-level-spacing distribution for
class AII$^\dag$ is calculated through identification of the two degenerate
eigenvalues.

\begin{figure*}[t]

\begin{center}
\includegraphics[width=\linewidth]{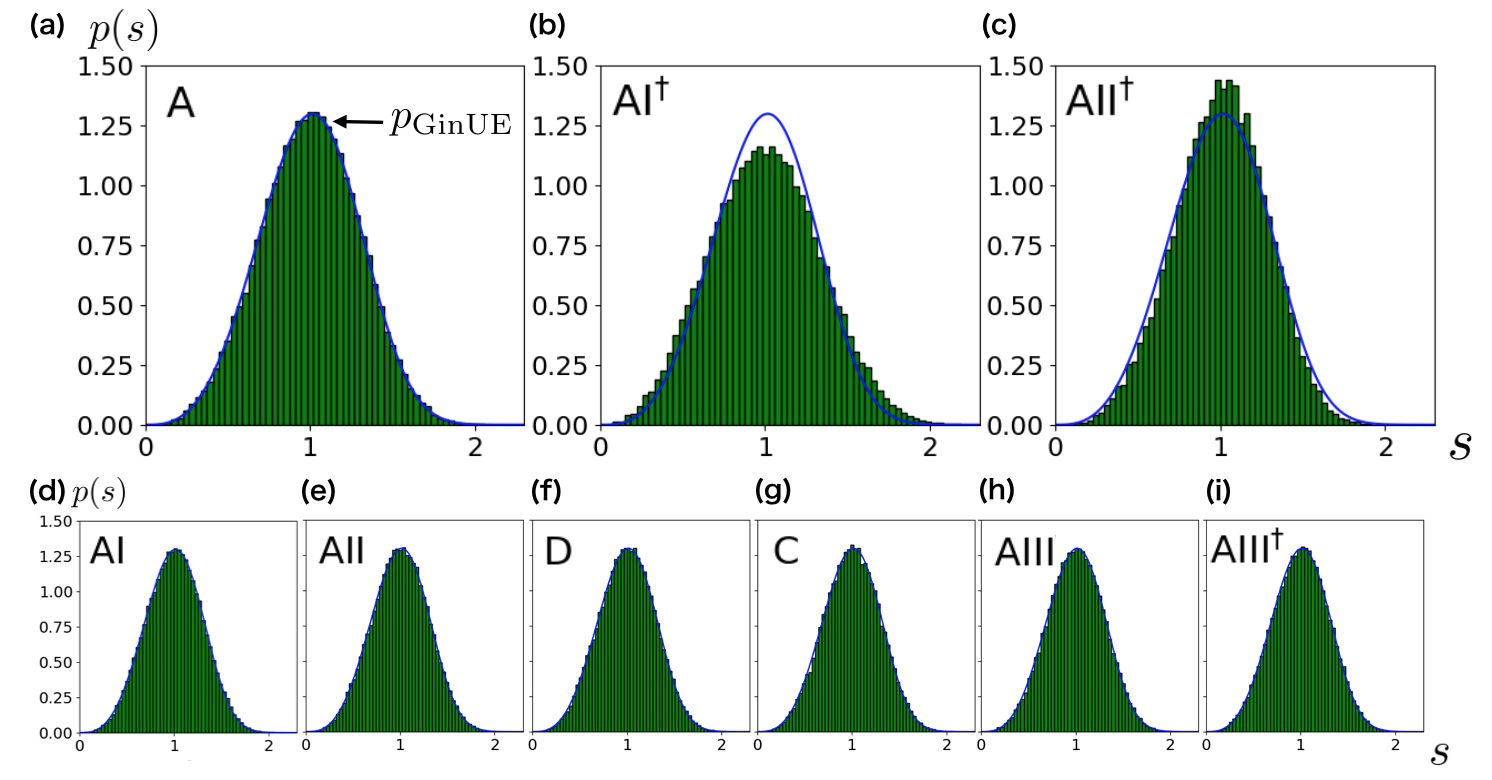}
\caption{Nearest-level-spacing distributions $p(s)$ for random matrices whose elements obey Gaussian distributions with their respective symmetry constraints.
The distributions $p(s)$ for classes (a) A, (d) AI, (e) D, (f) C, (g) AII, (h) AIII, and (i) AIII$^\dag$ obey the GinUE distribution $p_\mr{GinUE}(s)$.
In contrast, the peak of $p(s)$ is lower for class (b) AI$^\dag$  and higher for class (c) AII$^\dag$ compared with that of the GinUE.
These results are obtained from diagonalization of 6000 $\times$ 6000 matrices and averaging over 300 ensembles.
Statistics are taken from eigenvalues that
are away from the edges of the spectrum and the symmetric line (i.e., real  or imaginary axis) if it exists.
}
\label{fig1}
\end{center}
\end{figure*}

We conjecture that GinUE for symmetry classes A, AI, AII, D, C, AIII, and AIII$^\dag$ and the other two distributions for classes AI$^\dag$ and AII$^\dag$ are universal.
To further confirm that the universality appears for larger matrix sizes,
we show in Fig.~\ref{fig1} the results of $p(s)$ for 6000 $\times$ 6000 matrices for the nine different symmetry classes.
We find that classes A, AI, AII, D, C, AIII, and AIII$^\dag$ obey the distribution for GinUE, class AI$^\dag$ has a lower peak, and class AII$^\dag$ has a higher peak than that of the GinUE.
Similar results are obtained for the Bernoulli ensembles (data not shown).
We here note that the universality of class A and that of class AI have recently been rigorously proven under certain assumptions on the fourth moment of the random entries in the matrix elements~\cite{Tao15}.

\subsection{Cumulants of the spacing distributions}
\begin{figure*}[t]
\begin{center}
\includegraphics[width=\linewidth]{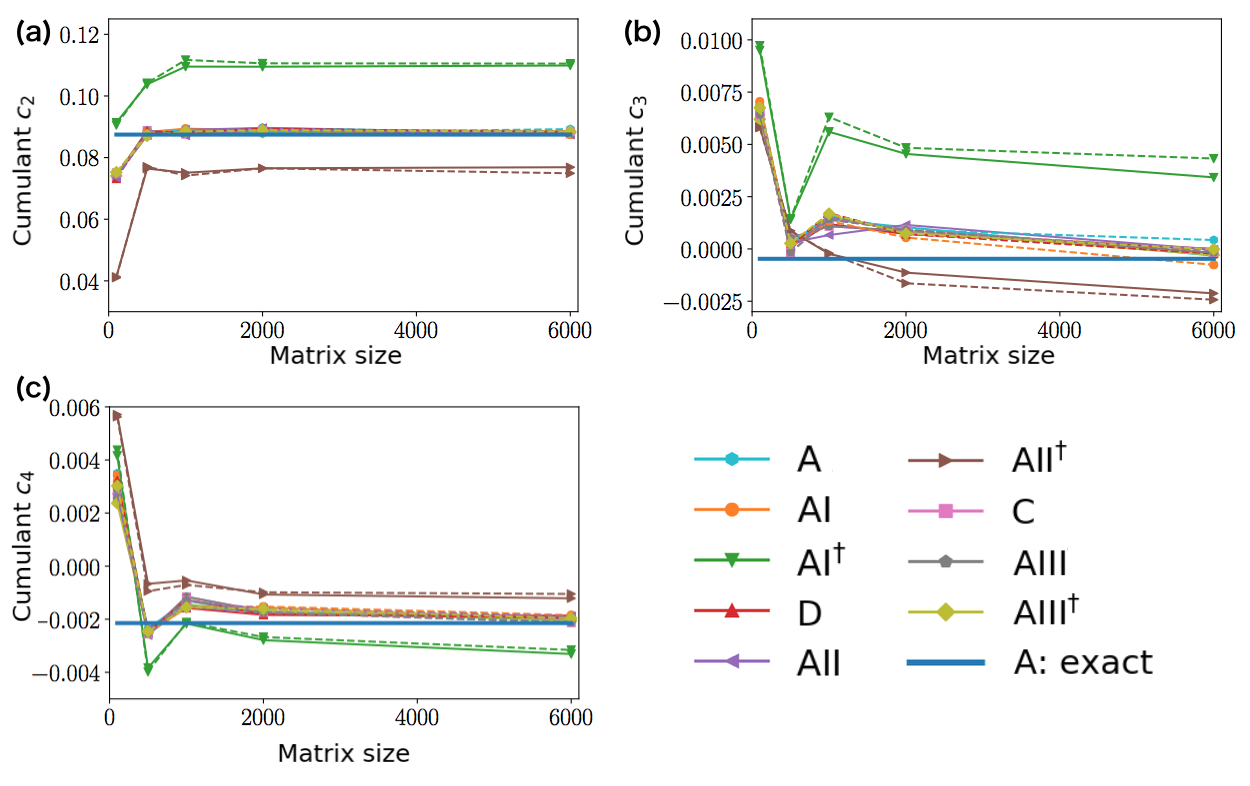}
\caption{
(a) Second, (b) third, and (c) fourth cumulants $c_2,c_3,c_4$ as a function of the matrix size for different symmetry classes and matrix ensembles (solid lines:  Gaussian ensembles, dashed lines:  Bernoulli ensembles).
The results for the Gaussian and Bernoulli ensembles of the same symmetry class are almost the same for all the cases.
For sufficiently large matrix sizes, we find three distinct universality classes even for the high-order cumulants;
classes A, AI, AII, D, C, AIII, and AIII$^\dag$ have the same cumulants that approach values calculated from the exact distribution $ p_\mr{GinUE}(s)$ (thick solid lines), but the cumulants for classes AI$^\dag$ and AII$^\dag$ approach different values.
Statistics are taken from eigenvalues that
are away from the edges of the spectrum and the symmetric line (real  or  imaginary axis) if it exists.
We average the data over 20000, 4000, 2000, 1000, 300 samples for the matrix  size of 100, 500, 1000, 2000, 6000, respectively.
}
\label{fig2}
\end{center}
\end{figure*}

Next, to confirm our universality more quantitatively, we study up to the fourth cumulants of $p(s)$ as a function of the size of a matrix for different symmetry classes and matrix distributions (Gaussian or Bernoulli).
The second, third, and fourth cumulants are given by
\aln{
c_2&=m_2-m_1^2,\\
c_3&=m_3-3m_1m_2+2m_1^3,\\
c_4&=m_4-4m_1m_3-3m_2^2+12m_1^2m_2-6m_1^4,
}
respectively,
where
\aln{
m_k=\int_0^\infty dss^kp(s)
}
is the $k$th moment.
We  fix the scale of $s$ such that  $m_1=1$.
As shown in Fig.~\ref{fig2}, we find the following properties that strengthen our argument on the universality:
\begin{itemize}
\item
The results for the Gaussian and Bernoulli ensembles in the same symmetry class are almost the same for all the cases.
Small deviations for the largest matrices are attributed to the limited number of samples used in our analysis.

\item
For sufficiently large matrix sizes, we clearly see three distinct universality classes even for the high-order cumulants:
classes A, AI, AII, D, C, AIII, and AIII$^\dag$ have the same cumulants, but classes AI$^\dag$ and AII$^\dag$ have different ones.
%Note that the relation for the second cumulant (or variance of the distribution), $ c_{\mr{AI^\dag},2}>c_{\mr{A},2}>c_{\mr{AII^\dag},2}$, is visibly understood from the

\item
The cumulants $c_{\mr{A},2}$, $c_{\mr{A},3}$, and $c_{\mr{A},4}$ approach values calculated from the exact distribution $ p_\mr{GinUE}(s)$ ($c_\mr{A,2}=0.0875,c_\mr{A,3}=-0.000471$, and $c_\mr{A,4}=-0.00215$).
However, the cumulants for classes AI$^\dag$ and AII$^\dag$ approach different values, which indicate that the newly found distributions for classes AI$^\dag$ and AII$^\dag$ are distinct from that of the GinUE even in the infinite-size limit.
From Fig.~\ref{fig2}, we conjecture that $c_{\mr{AI}^\dag,2}\simeq 0.11,c_{\mr{AII}^\dag,2}\simeq 0.075$,
 $c_{\mr{AI}^\dag,3}\simeq 0.004,c_{\mr{AII}^\dag,3}\simeq -0.0025$, and
  $c_{\mr{AI}^\dag,4}\simeq -0.003,c_{\mr{AII}^\dag,4}\simeq -0.001$.
\end{itemize}

These results suggest that only three universality classes for $p(s)$ exist.
Interestingly,
we find that $c_2$ (i.e., the variance) for class AI$^{\dag}$ (AII$^{\dag}$) is larger (smaller) than that for class A.
This behavior is similar to the Hermitian counterpart, where the variance for class AI (AII) is larger (smaller) than that for class A.
As discussed below, the broadening of the variance is attributed to an increase in the complex-valued degrees of freedom $f$ ($f=2, 3$, and $5$ for classes AI$^\dag$, A, and  AII$^\dag$, respectively) for the repulsive interaction between two neighboring eigenvalues.

We also  conjecture that only three universality classes due to TRS$^\dag$ exist for $p(s)$ among 38 non-Hermitian symmetry classes~\cite{Kawabata19S}, which may contain multiple symmetries.
Indeed, symmetries other than TRS$^\dag$ cannot alter  the local interaction between eigenvalues away from the symmetric lines.

\section{Analysis for small matrices}\label{sec4}
\subsection{Effective small matrices describing repulsive interactions}

 \begin{figure*}[t]
 \begin{center}
 \includegraphics[width=\linewidth]{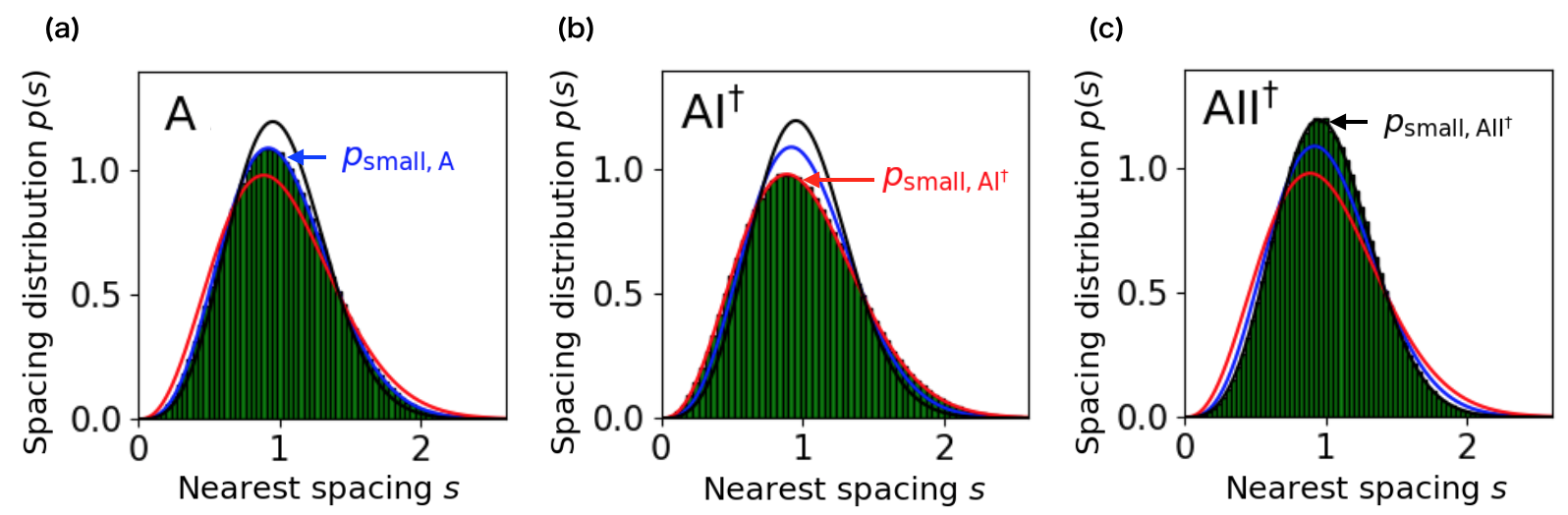}
 \caption{
 Nearest-level-spacing distributions $p(s)$ for two-by-two (for classes (a) A and (b) AI$^\dag$) and four-by-four (for (c) class AII$^\dag$) random matrices described by Eq.~(\ref{rmattwo}).
 The numerically obtained histograms are well described by analytical results in Eq.~(\ref{important}).
 Similar to Fig.~\ref{fig:fig2}(c-e), $p_\mr{\mr{small},AI^\dag}(s)$ (red curve) has a lower peak and $p_\mr{\mr{small},AII^\dag}(s)$ (black curve) has a higher peak than $p_\mr{\mr{small},A}(s)$ (blue curve).
 The results are obtained from averaging over $10^6$ ensembles.
 }
 \label{appfig3}
 \end{center}
 \end{figure*}

 The main feature of our results can be understood from the analysis of small matrices, i.e., the calculation of $p(s)$ obtained from two-by-two or four-by-four matrices.
 While the results of small matrices are quantitatively different from larger matrices~\cite{Haake}, we expect that it can describe some \textit{qualitative} features of level-spacing distributions, such as the change of the height and width of the distribution due to TRS$^\dag$~\cite{Grobe89}.
 This is because the repulsion between two close eigenvalues due to perturbations is qualitatively understood through diagonalization of the corresponding two-by-two or four-by-four transition matrix~\cite{Grobe89}.

 To see this, we consider a situation in which elements of a random matrix $H$ are slightly perturbed in a symmetry-preserving manner.
 The eigenvalues of the original matrix are correlated with this perturbation $V$.
 To simplify the problem, we assume that two (different) eigenvalues are much closer to each other than to the rest.
 Then the repulsion of these two eigenvalues can be estimated from diagonalization of $V$ in the subspace spanned by the corresponding eigenvectors.
 We expect that qualitative behavior of the nearest-level-spacing distributions can be described by this method~\cite{Grobe89,Haake}.

 The matrices that we obtain by the above method depend on the symmetry of $H$ (and equivalently $V$).
 Let us first consider the simplest case, class A, and let $\phi_1\:(\chi_1)$ and $\phi_2\:(\chi_2)$ be the corresponding right (left) eigenvectors.
 In this case, the corresponding two-by-two matrix is
 \aln{\label{smallA}
 \MATh{\chi_1^\dag V \phi_1}{\chi_1^\dag V \phi_2}{\chi_2^\dag V \phi_1}{\chi_2^\dag V \phi_2}=:\MATh{\mathfrak{a}}{\mathfrak{b}}{\mathfrak{c}}{\mathfrak{d}}.
 }
 Since $\phi_1$ and $\phi_2$ become independent random vectors for large $H$ and no direct relation between $\phi_\alpha$ and $\chi_\alpha$ exists in general, $\mathfrak{a, b, c, d}\in\mbb{C}$ can be regarded as independent random variables.
Let us assume that $\mathfrak{a, b, c,}$ and $\mathfrak{d}$ are  Gaussian random variables, motivated by the fact that the matrix elements with respect to random eigenvectors become Gaussian for the Hermitian cases~\cite{Brody81}.

  Next, we consider class AI.
  Whereas time-reversal symmetry imposes a constraint on two eigenvectors that are placed symmetrically around the real axis, it does not on two close eigenvectors that are off the real axis.
  Thus, the obtained matrix again has the form of Eq.~(\ref{smallA}).
  Note that $\mathfrak{a, b, c,}$ and $\mathfrak{d}$ are complex despite time-reversal symmetry of $V$, since the eigenvectors $\phi_1$ and $\phi_2$ spontaneously break time-reversal symmetry.
  In this sense, the interaction between two eigenvalues that are off the real or imaginary axis is also characterized by $H_\mr{small, A}$ even for class AI.
  The discussion also holds true for classes AII, D, C, AIII, and AIII$^\dag$:
  the eigenvectors $\phi_1$ and $\phi_2$ away from the real axis are regarded as two independent random vectors subject to no constraint, and we can consider $H_\mr{small, A}$.
  In other words, global reflection symmetry of the spectrum in the complex plane does not affect the local statistics away from the real or imaginary axis, as noted in Ref.~\cite{Haake}.

  On the other hand, the situation is different for classes with the TRS$^\dag$.
Here, we consider an arbitrary unitary matrix $\mc{C}_+$ (see Appendix~\ref{appA}) to clarify the generality of our discussion.
Note that we have chosen $\mc{C}_+=\mbb{I}_{N\times N}$ for class AI$^\dag$ and $\mc{C}_+=\Sigma^y$ for class AII$^\dag$ in the previous chapters.

  Let us consider class AI$^\dag$.
  In this case, we have a condition $\mc{C}_+\chi_\alpha^*= \phi_\alpha$ (we have chosen the proportionality factor to be unity).
  Then, while $\chi_1^\dag V \phi_1$ and $\chi_2^\dag V \phi_2$ become  independent complex variables, we obtain the following relation for the off-diagonal terms:
  \aln{
  \chi_1^\dag V \phi_2&=(\chi_1^\dag V \phi_2)^T\nonumber\\
  &=\phi_2^TV^T\chi_1^*\nonumber\\
  &=\chi_2^\dag \mc{C}_+^TV^T\mc{C}_+^{-1}\phi_1\nonumber\\
  &=\chi_2^\dag \mc{C}_+V^T\mc{C}_+^{-1}\phi_1\nonumber\\
  &=\chi_2^\dag V\phi_1,
  }
  where we have used $\mc{C}_+^T=\mc{C}_+$ for class AI$^\dag$.
  Thus, we obtain the symmetric matrix ($\mathfrak{b}=\mathfrak{c}$) in Eq.~(\ref{smallA}).
  %, which leads to $H_\mr{small, AI^\dag}$ in Eq. (1) in the main text.

  Finally, we consider class AII$^\dag$.
  Since $H$ has the Kramers degeneracy, we need to consider the four-by-four matrices spanned by $\phi_1$, $\av{\phi}_1$, $\phi_2$, and $\av{\phi}_2$ together with the corresponding left eigenvectors $\chi_1$, $\av{\chi}_1$, $\chi_2$, and $\av{\chi}_2$, where $\av{\phi}_\alpha=\mc{C}_+\chi_\alpha^*$ and $\av{\chi}_\alpha=(\phi_\alpha^T\mc{C}_+^{-1})^\dag$ are Kramers pairs.
  We have the following relation:
  \aln{
  \av{\chi}_\alpha^\dag V \av{\phi}_\beta&=\phi_\alpha^T\mc{C}_+^{-1} V\mc{C}_+\chi_\beta^*\nonumber\\
  &=(\phi_\alpha^T\mc{C}_+^{-1} V\mc{C}_+\chi_\beta^*)^T\nonumber\\
  &=\chi_\beta^\dag \mc{C}_+^TV^T(\mc{C}_+^T)^{-1}\phi_\alpha\nonumber\\
  &=\chi_\beta^\dag \mc{C}_+V^T\mc{C}_+^{-1}\phi_\alpha\nonumber\\
  &=\chi_\beta^\dag V\phi_\alpha,
  }
  where we have used $\mc{C}_+^T=-\mc{C}_+$ for class AII$^\dag$.
  We also have
  \aln{
  {\chi}_\alpha^\dag V \av{\phi}_\beta&=({\chi}_\alpha^\dag V \mc{C}_+{\chi}_\beta^*)^T\nonumber\\
  &=\chi_\beta^\dag \mc{C}_+^T V^T\chi_\alpha^*\nonumber\\
  &=-\chi_\beta^\dag \mc{C}_+ V^T\chi_\alpha^*\nonumber\\
  &=-\chi_\beta^\dag V\mc{C}_+ \chi_\alpha^*\nonumber\\
  &=-\chi_\beta^\dag V\av{\phi}_\alpha
  }
  and
  \aln{
  \av{\chi}_\alpha^\dag V {\phi}_\beta&=({\phi}_\alpha^T\mc{C}_+^{-1} V\phi_\beta)^T\nonumber\\
  &=\phi_\beta^T V^T(\mc{C}_+^T)^{-1} \phi_\alpha\nonumber\\
  &=-\phi_\beta^T V^T\mc{C}_+^{-1} \phi_\alpha\nonumber\\
  &=-\phi_\beta^T \mc{C}_+^{-1}V \phi_\alpha\nonumber\\
  &=-\av{\chi}_\beta^\dag V \phi_\alpha.
  }
%  Combining them, we obtain $H_\mr{small, AII^\dag}$ in Eq. (1) in the main text.

We can now analyze
 \aln{\label{rmattwo}
 H_{\mr{small}, \mr{A}}&=\MATh{\mathfrak{a}}{\mathfrak{b}}{\mathfrak{c}}{\mathfrak{d}},\:\:\:
 H_{\mr{small}, \mr{AI}^\dag}=\MATh{\mathfrak{a}}{\mathfrak{b}}{\mathfrak{b}}{\mathfrak{d}},\:\:\:\nonumber\\
 H_{\mr{small}, \mr{AII}^\dag}&=
 \begin{pmatrix}
 \mathfrak{a} & 0 & \mathfrak{e}&\mathfrak{b}\\
 0 & \mathfrak{a} & \mathfrak{c}&\mathfrak{g}\\
 \mathfrak{g} & -\mathfrak{b} & \mathfrak{d}&0\\
 -\mathfrak{c} & \mathfrak{e} & 0&\mathfrak{d}\\
 \end{pmatrix}
 }
 for $\mathfrak{a, b, c, d ,e, g}\in\mbb{C}$.
 Note that $  H_{\mr{small}, \mr{A}}$ also describes classes AI, AII, D, C, AIII, and AIII$^\dag$.
 Notably, as \textit{complex} variables, the degrees of freedom determining the level-spacing distribution are $f=2$ for class AI$^\dag$, 3 for class A, and 5 for class AII$^\dag$ (see Table~\ref{Table threefold ways}). Here, one degree is dropped from all the
 variables because the trace of the matrix does not affect the level
 spacings.
 This is a natural generalization of the Hermitian case, where $f=2$ for  class AI, 3 for class A, and class 5 for AII as \textit{real} variables~\cite{Haake}.

 As shown in Appendix~\ref{appB}, the analytic expressions of the level-spacing distributions $p_{\mr{small}}$ with the complex-valued degrees of freedom $f$ for Gaussian-distributed small matrices are obtained as
\begin{align}\label{important}
 p_\mr{small}(s) = \frac{(C_f s)^3}{\mc{N}_f}  K_{\frac{f-2}{2}}\lrs{(C_f s)^2},
 \end{align}
where $K_\nu(x)=\int_{0}^{\infty}dz e^{-x\cosh z}\cosh(\nu z)$ is the modified Bessel function, and
 $C_f$  and $\mc{N}_f$
 are some constants.
 The forms of $p(s)$ for specific values of $f$ can also be given by
 \aln{\label{important2}
  p_\mr{\mr{small},A}(s)&=2C_3^4s^3e^{-C_3^2s^2},\nonumber\\
  p_\mr{\mr{small},AI^\dag}(s)&=2{C_2^4}s^3K_0\lrs{{C_2^2s^2}},\nonumber\\
  p_\mr{\mr{small}, AII^\dag}(s)&=\frac{2C_5^4s^3}{3}\lrs{1+{C_5^2s^2}}e^{-{C_5^2s^2}},
 }
 where $C_2=\frac{1}{8\sqrt{2}}\Gamma\lrs{\frac{1}{4}}^2=1.16187\dots$,
 $C_3=\frac{3}{4}\sqrt{\pi}=1.32934\dots$,
  and $C_5=\frac{7}{8}\sqrt{\pi}\simeq 1.5509\dots$.
 Here
 \aln{
 K_{1/2}(x)&=\sqrt{\frac{\pi}{2x}}e^{-x}, \\
 K_{3/2}(x)&=\sqrt{\frac{\pi}{2x}}\lrs{1+\frac{1}{x}}e^{-x}
 }
are used to obtain Eq.~\eqref{important2}.
 In Fig.~\ref{appfig3}, we plot Eq.~\eqref{important} (or Eq.~\eqref{important2}) and compare it with numerical results, which shows a perfect agreement.
 Note that, in a manner similar to Fig.~\ref{fig:fig2}(c-e), $p_\mr{\mr{small},AI^\dag}(s)$ has a lower peak and $p_\mr{\mr{small},AII^\dag}(s)$ has a higher peak compared with $p_\mr{\mr{small},A}(s)$.

 The analytic forms in Eq.~\eqref{important} are derived in a unified manner:
 $p_\mr{small}(s)$ is understood as the distribution of \aln{
 s=\sqrt{\lrv{\sum_{i=1}^fz_f^2}},}
  where $z_i$ are complex Gaussian random variables.
All of the three distributions are then obtained from transformations of the so-called $K$-distribution with a different shape parameter for each class~\cite{Jakeman78} (see Appendix~\ref{appB} for details).
 We also perform a similar analysis for Hermitian small matrices:  $p_\mr{small}(s)$ is understood by the distribution of $s=\sqrt{\sum_{i=1}^fx_f^2}$ with real Gaussian random variables $x_i$, which leads to simple transformations of the chi-squared distributions (see Table~\ref{Table threefold ways}).
  To our best knowledge, such a unified interpretation of level-spacing distributions for Hermitian matrices from the chi-squared distribution has never been presented.

  In the {H}ermitian case, the asymptotic
  behavior of chi-squared distributions leads to the class-dependent
  level-repulsion factor. However, the $K$-distribution has the same asymptotic
  behavior except for the logarithmic correction in class AI$^\dag$, leading to
  the class-independent factor $s^3$ as noted in Ref.~\cite{Grobe89} for the
  non-{H}ermitian case.

 While the level-repulsion factor of $p_\mr{small}(s\ra 0)$ is universally $\sim s^3$ in the non-Hermitian case, the entire distribution depends on the symmetry class  as in the Hermitian case, since  TRS$^\dag$ alters the repulsive interactions between complex eigenvalues through $f$.
 Indeed, the peak  of $p_\mr{\mr{small},AI^\dag}(s)$ is lower and that of $p_\mr{\mr{small},AII^\dag}(s)$ is higher than that of $p_\mr{\mr{small},A}(s)$ in Eq.~\eqref{important},
and the normalized variances $v_{\mr{small}}$ satisfy (see Fig.~\ref{appfig3})
\aln{
v_{\mr{small},\mr{AI}^\dag}>v_{\mr{small},\mr{A}}>v_{\mr{small},\mr{AII}^\dag}.
}
We can intuitively understand this behavior from the fact that $p_\mr{small}(s)$ is  the normalized distribution of $s=\sqrt{\lrv{\sum_{i=1}^fz_f^2}}$, whose variance becomes smaller for larger $f$.

 The above properties for the peak and variance also hold true in numerical calculations of large matrices as seen from suppression or enhancement of the Ginibre distribution in Fig.~\ref{fig:fig2}.
%As detailed in the next section, such changes due to transposition symmetry account for the distinctive behavior of $p(s)$ in open quantum  systems, which allows for unambiguous characterization of nonintegrability and chaos in dissipative systems with different symmetry.
The detailed pieces of information such as the peak and variance are needed to fully characterize chaos even in Hermitian systems~\cite{Bogomolny99,Garcia06}.
We also note that the relation between cumulants higher than the second is not necessarily revealed by small matrices, since we ignore the effect of nearby eigenvalues other than two close ones in calculating $p_\mr{small}(s)$.

\begin{table}[tb]
 \caption{\label{Table threefold ways}
  Wigner-Dyson's Hermitian three universal statistics with TRS and
   our non-Hermitian three universal statistics with TRS$^{\dag}$ for Gaussian-distributed small matrices.
   While they share the same degrees of freedom $f=2,3$, and $5$,
   the distinction between real-valued and complex-valued degrees leads to the different families of
   level-spacing distributions, i.e., the transformed chi-squared distribution for the Hermitian case and the  $K$-distribution for the non-Hermitian case.
  }

 \centering
 \begin{tabular}{ccc}
  \hline \hline
  & Hermitian (TRS) & Non-Hermitian (TRS$^\dag$) \\
  \hline
  $f$ & $2\:(\mr{AI}),3\:(\mr{A}),5\:(\mr{AII})$ & $2\:(\mr{AI}^\dag),3\:(\mr{A}),5\:(\mr{AII}^\dag)$ \\
  & (as real variables) & (as complex variables) \\
  \hline
  $p_\mr{small}$  & chi-squared distribution & $K$-distribution \\
   $v_\mr{small}$ &
      AI $>$ A $>$ AII& AI$^\dag$ $>$ A $>$ AII$^\dag$ \\
  \hline \hline
 \end{tabular}
\end{table}

\begin{figure*}[t]
\begin{center}
\includegraphics[width=\linewidth]{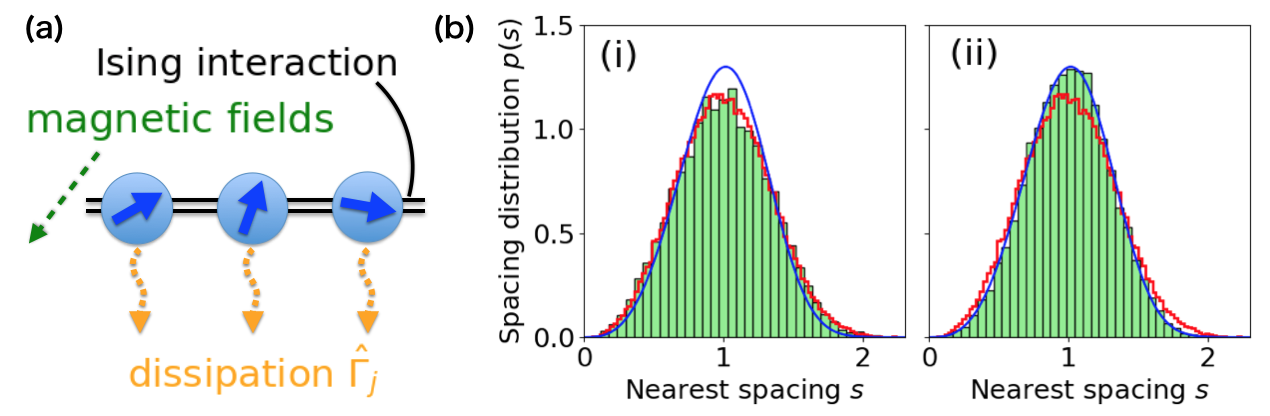}
\caption{
(a) Schematic illustration of a locally interacting system described by the Lindblad equation~\eqref{siki2}.
(b) Nearest-level-spacing distributions $p(s)$ for two different dissipation mechanisms [(i) dephasing $\hat{\sigma}^z_j$ and (ii) damping $\hat{\sigma}^-_j$] after the unfolding procedure of supereigenvalues.
Comparison with Fig.~\ref{fig:fig2}(c-e) reveals that local correlations of the Liouvillian spectra are well described by non-Hermitian random matrices with their respective symmetry:
model (i) belongs to class AI$^\dag$ universality (red curve, the same data in Fig~\ref{fig:fig2}(d), while model (ii) belongs to class A universality (blue curve).
The results are obtained from averaging over 10 ensembles for $\gamma=0.75$ and $L=7$.
Statistics are taken from supereigenvalues that
are away from the edges of the spectrum.
}
\label{fig:fig4}
\end{center}
\end{figure*}

\section{Universality in dissipative quantum many-body systems}\label{sec5}
As a direct application of the new universality for the level-spacing distributions, we discuss the classification of dissipative quantum many-body chaos.
Other statistics, such as the statistics of the normalized largest eigenvalues of random {H}ermitian matrices (i.e., the Tracy-Widom distribution) are relevant to other physics such as fluctuations of a growing  interface~\cite{Takeuchi10}.
On the other hand, the characterization of chaotic systems requires universal local statistics in the bulk of the spectrum, represented by the level-spacing distributions.

It is conjectured~\cite{Bohigas84} that the level-spacing statistics for Hermitian quantum chaotic Hamiltonians obey those of random matrices~\cite{Berry77R,Bohigas84,Peres84E2,Feingold86,Srednicki94,Srednicki99,Muller04,Rigol08,Santos10a,Pal10,Khatami13,Beugeling14,Hamazaki16G,Beugeling15,DAlessio16,Luitz16,Mondaini16,Kos17,Mondaini17,Hamazaki18A}, in contrast with integrable systems that obey the Poisson distributions.
As a natural generalization of this conjecture, the level-spacing statistics of non-Hermitian random matrices have been applied to probe the nonintegrability-integrability transition and the chaos-localization transition in open quantum systems~\cite{Grobe88,Markum99,Hamazaki19N}.
However, it has remained unclear how the distributions change with symmetry.
We here demonstrate that the newly found universality classes appear in dissipative quantum many-body systems described by a Lindblad equation and a non-Hermitian Hamiltonian.
These results show that the newly found universality classes play a key role in characterizing nonintegraiblity of dissipative many-body systems.

\subsection{Lindblad many-body equation}
We introduce a dissipative one-dimensional spin-1/2 model described by the Lindblad equation $\fracd{\hrho}{t}=\mc{L}[\hrho]$ (Fig~\ref{fig:fig4}(a))~\cite{Breuer02}, where
\aln{\label{siki2}
 \mc{L}[\hrho]
 =-i[\hat{H}, \hat{\rho}]+\sum_{j=1}^L \gamma \lrl{\hat{\Gamma}_{j}\hat{\rho}\hat{\Gamma}_{j}^\dag-\frac{1}{2}\lrm{\hat{\Gamma}_{j}^\dag\hat{\Gamma}_{j},\hrho}}.
}
The Hamiltonian is an Ising model with transverse and longitudinal fields
\aln{
 \hat { H }  = - \sum _ { j = 1 } ^ { L - 1 }(1+ \epsilon _ { j }) \hat { \sigma } _ { j } ^ { z } \hat { \sigma } _ { j + 1 } ^ { z } - \sum _ { j = 1 } ^ { L } \lrs{ -1.05 \hat { \sigma } _ { j } ^ { x } +  0.2\hat { \sigma } _ { j } ^ { z } }
}
with $\epsilon_j$ randomly chosen from $[-0.1,0.1]$ for each site $j$ to break unwanted symmetries.
Local dissipations $\hat{\Gamma}_j$ are either (i) dephasing $\hat{\sigma}^z_j$ or (ii) damping $\hat{\sigma}^-_j$.
This model can be realized with Rydberg atoms~\cite{Lee12}.

We consider the Liouvillian spectrum of the super-operator $\mc{L}$ in Eq.~\eqref{siki2}.
Namely, we consider the  super-eigenvalue equation
\aln{
\mc{L}[\hat{\nu}_\alpha]=\lambda_\alpha\hat{\nu}_\alpha,
}
where $\lambda_\alpha$ is a super-eigenvalue for a super-eigenstate $\hat{\nu}_\alpha$.

When we consider $\mc{L}$ as a matrix,  model (i) has a transposition symmetry ($\hat{\sigma}_j^{x,z}\ra\hat{\sigma}_j^{x,z}, \hat{\sigma}_j^{y}\ra-\hat{\sigma}_j^{y}$) but  model (ii) does not.
Both $\mc{L}$ have additional TRS, but it does not affect the level-spacing distribution away from the real axis as discussed above.

To understand this symmetry structure, it is convenient
to use the operator representation of super-operators.
That is, we consider the isomorphic mapping
\aln{
\hat{A}\ket{i}\bra{j}\hat{B}\ra(\hat{A}\otimes\hat{B}^T)\ket{i}\otimes\ket{j},
}
where we have doubled the Hilbert space by adding a dual space.
Then, the Lindblad super-operator can be represented  by
\aln{
\mc{L}\ra\hat{\mc{L}}&=-i(\hH\otimes \mbb{I}-\hat{\mbb{I}}\otimes\hH^T)\nonumber\\
&+\gamma\sum_{j=1}^L\lrl{\hat{\Gamma}_j\otimes\hat{\Gamma}^*
_j-\frac{1}{2}\hat{\Gamma}_j^\dag\hat{\Gamma}_j\otimes \hat{\mbb{I}}
-\hat{\mbb{I}}\otimes \hat{\Gamma}^T_j\hat{\Gamma}^*_j},
}
whose eigenvalues are $\lambda_\alpha$.

For arbitrary $\hat{\Gamma}_j$, $\hat{\mc{L}}$ has TRS with $\mathcal{T}_{+} \mathcal{T}_{+}^{*} = +1$.
Indeed, for the unitary swap operation $\mc{T}_+=\mbb{SWAP}$ that exchanges the original and the copied Hilbert spaces (i.e., $\mc{T}_+(\hat{A}\otimes\hat{B})\mc{T}_+^{-1}=\hat{B}\otimes\hat{A}$), $\mc{T}_+\mc{\hat{L}}^*\mc{T}_+^{-1}=\mc{\hat{L}} $ and $ \mc{T}_+\mc{T}_+^*=+1$ (see Eq.~\eqref{eq: app I - AI}) are satisfied because $\hat{H}=\hat{H}^\dag$.

On the other hand, for our Ising model with transverse and longitudinal fields, $\mc{\hat{L}}$ can also have TRS$^\dag$ depending on $\hat{\Gamma}_j$.
Indeed, since $\hat{H}=\hat{H}^T$ in the conventional Pauli basis, we have
\aln{
\hat{\mc{L}}^T&=-i(\hH\otimes \hat{ \mbb{I}}-\hat{\mbb{I}}\otimes\hH^T)\nonumber\\
&+\gamma\sum_{j=1}^L\lrl{\hat{\Gamma}_j^T\otimes\hat{\Gamma}^\dag
_j-\frac{1}{2}\hat{\Gamma}_j^T\hat{\Gamma}_j^*\otimes \hat{\mbb{I}}
-\hat{\mbb{I}}\otimes \hat{\Gamma}_j^\dag\hat{\Gamma}_j}.
}
Thus, if
\aln{
\hat{\Gamma}_j^T=\hat{\Gamma}_j,\:\:\:
%(\hat{\Gamma}_j^\dag\hat{\Gamma}_j)^T=\hat{\Gamma}_j^\dag\hat{\Gamma}_j,
[\hat{\Gamma}_j^\dag,\hat{\Gamma}_j]=0,
}
we have the transposition symmetry, $\mc{\hat{L}}^T=\mc{\hat{L}}$.
This condition is satisfied for (i) dephasing $ \hat{\Gamma}_j=\hat{\sigma}_j^z$ but not for (ii) damping $ \hat{\Gamma}_j=\hat{\sigma}_j^-$.

Figure~\ref{fig:fig4}(b) shows $p(s)$ for the two models (i) and (ii) after the unfolding procedure of the spectra.
We can clearly see that there appear distinct distributions that correspond to the universality classes of the random-matrix ensembles in class AI$^\dag$ or A in Fig.~\ref{fig:fig2}.
Note that the model in Eq.~\eqref{siki2} generally exhibits properties which are very different from those predicted by random matrices in that the matrix is sparse due to the local interactions and that the randomness $\epsilon_j$ is small.
Nevertheless, our results show that local correlations of eigenenergies of nonintegrable dissipative Lindblad systems are well described by non-Hermitian random matrices that take into account transposition symmetry.

\subsection{Non-Hermtian many-body Hamiltonian}
\begin{figure*}[t]
\begin{center}
\includegraphics[width=\linewidth]{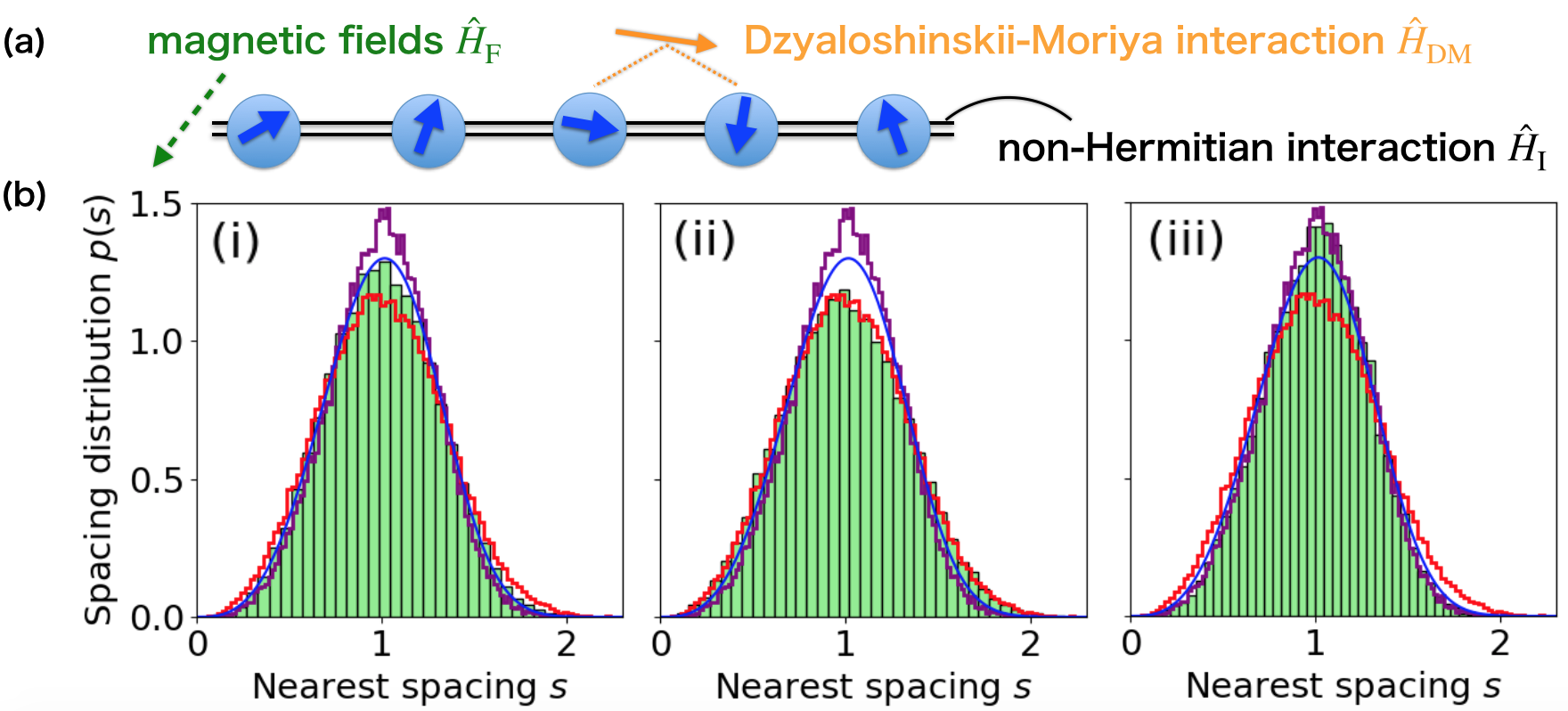}
\caption{
(a) Schematic illustration of a locally interacting spin system described by the non-Hermitian Hamiltonian~\eqref{siki}.
(b) Nearest-level-spacing distributions $p(s)$ for the three non-Hermitian models (i), (ii), and (iii) in Eq.~\eqref{siki} after the unfolding procedure of eigenenergies.
Comparison with Fig.~\ref{fig:fig2}  reveals that local correlations of eigenenergies of the dissipative systems are well described by non-Hermitian random matrices with their respective symmetry:
model (i) belongs to class A (blue curve), model (ii) belongs class AI$^\dag$ (red curve, the same data with Fig.~\ref{fig:fig2}(d)), and model (iii) belongs class AII$^\dag$ (purple curve, the same data with Fig.~\ref{fig:fig2}(e)).
The parameters used are $h=0.5, D=0$ for model (i), $h=0.5, D=0.9$ for model (ii), and $h=0, D=0.9$ for model (iii), and $J=0.2$ and $L=13$ for all the models.
The results are obtained from averaging over 30 ensembles.
Statistics are taken from eigenenergies that
are away from the edges of the spectrum.
}
\label{fig:fig42}
\end{center}
\end{figure*}

We also find the universal results for non-Hermitian many-body systems which belong to class AII$^\dag$ as well as A or AI$^\dag$.
We consider a one-dimensional spin-1/2 model with a non-Hermitian Ising interaction, transverse and longitudinal fields, and the  Dzyaloshinskii-Moriya interaction as follows (Fig~\ref{fig:fig42}(a)):
\aln{\label{siki}
 \hat { H }(J,h,D) & = \hat { H } _ \mr{ I }(J) + \hat { H } _ \mr{ F }(h) + \hat { H } _ \mr{ D M }(D) ,
}
where
\aln{
 \hat { H } _ { \mathrm { I } }(J) & = - \sum _ { j = 1 } ^ { L - 1 }(1+ iJ\epsilon _ { j }) \hat { \sigma } _ { j } ^ { z } \hat { \sigma } _ { j + 1 } ^ { z } ,\nonumber \\
 \hat { H } _ { \mathrm { F } }(h) & = - h\sum _ { j = 1 } ^ { L } \lrs{ -2.1 \hat { \sigma } _ { j } ^ { x } +  \hat { \sigma } _ { j } ^ { z } }, \nonumber\\
 \hat { H } _ { \mathrm { DM } }(D)& =\sum _ { j = 1 } ^ { L - 1 } \vec { D } \cdot ( \vec { \hat { \sigma } } _ { j } \times \vec { \hat { \sigma } } _ { j + 1 } ).
 }
Here, $L$ is the size of the system, $\epsilon_j$ is chosen randomly from $[-1,1]$ for each site $j$ to break unwanted symmetries, $\vec { D } := \frac { D } { \sqrt { 2 } } \left( \vec { e } _ { x } + \vec { e } _ { z } \right)$, and the open boundary condition is assumed.
The numerical factor 2.1 in $\hat{H}_\mr{F}$ is chosen such that the model becomes sufficiently nonintegrable.
Note that non-Hermitian many-body systems can be realized in continuously measured quantum many-body systems~\cite{Daley14,Lee14}.
In particular, the non-Hermitian term in Eq.~\eqref{siki} arises if we consider the collective dephasing  $\hat{\sigma}^z_j\hat{\sigma}^z_{j+1}$ and postselect the null measurement outcome.

The symmetry of this model depends on the parameters $J, h$, and $D$.
For $J\neq 0$, the model belongs to  (i) class A for $h\neq 0, D\neq 0$, (ii) class AI$^\dag$ for $h\neq 0, D=0$ because $\hat{H}=\hat{H}^T$, and (iii) class AII$^\dag$ for  $h=0,D\neq 0$ and odd $L$ because $\hat{H}=\lrs{\prod_{i=1}^{L}\hat{\sigma}_i^y}\hat{H}^T\lrs{\prod_{i=1}^{L}\hat{\sigma}_i^y}^{-1}$ with $\lrs{\prod_{i=1}^{L}\hat{\sigma}_i^y}\lrs{\prod_{i=1}^{L}\hat{\sigma}_i^y}^*=(-1)^L$.
Note that, for $J=0$, the model reduces to ({H}ermitian) classes (i) A, (ii) AI, and (iii) AII, whose level-spacing distributions become (i) GUE, (ii) GOE, and (iii) GSE, respectively~\cite{Hamazaki19R,HamazakiM}.

Figure~\ref{fig:fig42}(b) shows the distributions $p(s)$ for the non-Hermitian models (i), (ii), and (iii) after the unfolding procedure of eigenenergies~\cite{Haake}.
We can clearly see that there appear distinct distributions that correspond to the universality classes of the random-matrix ensembles with the same symmetries as  in Fig.~\ref{fig:fig2}.
Note that the model in Eq.~\eqref{siki} generally exhibits properties very different from the predictions of random matrices in that the matrix is sparse due to the local interactions and that the randomness $J\epsilon_j$ is small.
Nevertheless, our results show that local correlations of eigenenergies of nonintegrable non-Hermitian systems are well described by non-Hermitian random matrices with the corresponding symmetry.

\section{Conclusion}\label{sec6}
We discover two new universality classes distinct from the GinUE exclusively for symmetry classes AI$^\dag$ and AII$^\dag$ by investigating the universality of the nearest-neighbor spacing distributions of non-Hermitian random matrices.
The three universal statistics for classes A, AI$^\dag$, and AII$^\dag$ defined by TRS$^\dag$ (transposition symmetry) constitute a natural generalization of Wigner-Dyson's  universal statistics for Hermitian matrices.
Only three universality classes due to TRS$^\dag$ are expected to exist for the level-spacing distribution away from the symmetric lines among 38 non-Hermitian symmetry classes~\cite{Kawabata19S}.
Our results serve as a basis for characterizing nonintegrability and chaos in open quantum systems with symmetry.

Our work paves the way toward understanding universality in non-Hermitian systems and raises many open questions.
It merits further study to investigate if a new universality appears for other statistics, such as correlation functions of distant eigenvalues and  distributions of the edges of the spectrum.
We note that TRS$^\dag$ appears in a wide range of physical systems in addition to dissipative many-body systems reported here, such as  systems with gain and loss~\cite{ElGanainy18,Miri19}, and classical many-body dynamics for a set of coupled oscillators~\cite{Huber16}.
It is an interesting future problem to investigate how our new classes lead to universal physical phenomena (such as nonintegrablity-integrability transitions) in those systems, as Dyson's classes have done in Hermitian systems.

\textit{Note added.}
 After our manuscript appears on arXiv, two related works appeared~\cite{Akemann19,Sa19}, which discuss the appearance of non-Hermitian random-matrix universality in dissipative quantum many-body systems.

\begin{acknowledgments}
We thank Masatoshi Sato, Hideaki Obuse, and Masaki Sano for helpful discussions.
This work was supported by
KAKENHI Grant No. JP18H01145 and
a Grant-in-Aid for Scientific Research on Innovative Areas ``Topological Materials Science" (KAKENHI Grant No. JP15H05855)
from the Japan Society for the Promotion of Science (JSPS).
R.H., K.K., and N.K. were supported by the JSPS through the Program for Leading Graduate Schools (ALPS) and KAKENHI Grant Nos. JP17J03189, JP19J21927, and JP19J11303.
\end{acknowledgments}

\appendix

\section{Symmetry classes and their fundamental properties}
~\label{appA}
We formulate symmetries in non-Hermitian systems and describe their fundamental properties according to the classification in Ref.~\cite{Kawabata19S}. We also derive explicit forms of the probability distributions of Gaussian ensembles and Bernoulli ensembles for each symmetry class. In the following, a complex eigenvalue is denoted as $E_{\alpha}$ and the corresponding right (left) eigenvector is denoted as $\phi_{\alpha}$ ($\chi_{\alpha}$):
\begin{equation}
H \phi_{\alpha} = E_{\alpha} \phi_{\alpha},\quad
\chi_{\alpha}^{\dag} H = \chi_{\alpha}^{\dag} E_{\alpha}.
	\label{eq: app I - eigenequation}
\end{equation}

\subsection{Symmetry for non-Hermitian matrices}

Non-Hermiticity alters the nature of symmetry in a fundamental manner. In particular, non-Hermiticity ramifies symmetry~\cite{Kawabata19S}. To see this symmetry ramification, let us consider time-reversal symmetry (TRS) as an example. For Hermitian matrices, TRS is defined by
$
{\cal T}\,H^{*}\,{\cal T}^{-1} = H,
 $
where ${\cal T}$ is a unitary matrix.
We can similarly define TRS for non-Hermitian matrices as
\aln{
{\cal T_+}\,H^{*}\,{\cal T}_+^{-1} = H,
	\label{eq: PHS cc}
}
where $\mc{T}_+$ is a unitary matrix.
However, for non-Hermitian matrices, we can define TRS in a different way. The crucial observation here is that complex conjugation is equivalent to  transposition for Hermitian matrices: $H^{*} = H^{T}$. As a result, ${\cal T}\,H^{*}\,{\cal T}^{-1} = H$ is equivalent to
$ {\cal T}\,H^{T}\,{\cal T}^{-1} = H$ for Hermitian $H$.
However, since $H^{*} \neq H^{T}$ for non-Hermitian matrices,  we can define another symmetry $\mc{C}_+$ as
\aln{
{\cal C_+}\,H^{T}\,{\cal C}_+^{-1} = H,
	\label{eq: PHS tt}
}
where $\mc{C}_+$ is a unitary matrix which is different from $\mc{T}_+$.
 Since the physical TRS is described by Eq.~(\ref{eq: PHS cc})~\cite{Kawabata19S}, we refer to the symmetry in Eq.~(\ref{eq: PHS cc}) as TRS for non-Hermitian matrices, and the symmetry in Eq.~(\ref{eq: PHS tt}) as TRS$^\dag$.

Such symmetry ramification occurs also for all the other symmetries. For example, let us consider chiral symmetry (CS), which is defined for Hermitian matrices by
\begin{equation}
\Gamma\,H^\dag\,\Gamma^{-1} = - H,
	\label{eq: CS no}
\end{equation}
where $\Gamma$ is a unitary matrix.
Note that CS is equivalent to sublattice symmetry (SLS) for Hermitian matrices, which is defined without Hermitian conjugation (i.e., $\Gamma H \Gamma = - H$).
Equation~(\ref{eq: CS no}) can be directly generalized to non-Hermitian matrices,
but again, CS can be generalized in a different manner.
In fact, for non-Hermitian matrices, Eq.~(\ref{eq: CS no}) is different from sublattice symmetry, defined by
\begin{equation}
\mc{S}\,H\,\mc{S}^{-1} = - H
	\label{eq: CS dag}
\end{equation}
because $H \neq H^{\dag}$.
Since the physical CS is described by Eq.~(\ref{eq: CS no})~\cite{Kawabata19S}, we refer to the symmetry in Eq.~(\ref{eq: CS no}) as CS, and the symmetry defined by Eq.~(\ref{eq: CS no}) as $\mathrm{CS}^{\dag}$ or SLS for non-Hermitian matrices.
%To distinguish $\Gamma$ of CS non-Hermitian matrices, we use $\mc{S}$ to describe SLS for non-Hermitian matrices.

In a similar manner, particle-hole symmetry (PHS) for Hermitian matrices, satisfying $\mc{C}{H}^T\mc{C}^{-1}=\mc{C}{H}^*\mc{C}^{-1}=-H$,
 ramifies in the presence of non-Hermiticity.
In the main text and the following discussions, we define PHS for non-Hermitian matrices by $\mc{C}_-{H}^T\mc{C}_-^{-1}=-H$ together with a unitary matrix $\mathcal{C}_{-}$.
On the other hand, we define PHS$^\dag$ for non-Hermitian matrices by $\mc{T}_-{H}^*\mc{T}_-^{-1}=-H$ together with a unitary matrix $\mathcal{T}_{-}$.

Non-Hermiticity not only ramifies but also unifies symmetry~\cite{Kawabata18T}. To see this symmetry unification, we consider the following antiunitary symmetries:
\begin{equation}
{\cal T}_{+}\,H^{*}\,{\cal T}_{+}^{-1} = H,\quad
{\cal T}_{-}\,H^{*}\,{\cal T}_{-}^{-1} = - H.
\end{equation}
Here ${\cal T}_{+}$ denotes TRS, while ${\cal T}_{-}$ denotes ${\rm PHS}^{\dag}$. TRS and PHS are clearly distinct from each other for Hermitian matrices. However, when a non-Hermitian matrix $H$ respects TRS, the non-Hermitian matrix $iH$ respects ${\rm PHS}^{\dag}$. Thus, a set of all the non-Hermitian matrices with TRS coincides with a different set of all the non-Hermitian matrices with ${\rm PHS}^{\dag}$; non-Hermiticity unifies TRS and ${\rm PHS}^{\dag}$.

As a result of the symmetry ramification, the 5 classes (AIII, AI, D, AII, and C)  for Hermitian matrices with a  single symmetry (TRS, PHS, or CS) bifurcate into the 10 classes  (AIII, AI, D, AII, C, AIII$^\dag$, AI$^\dag$, D$^\dag$, AII$^\dag$, and C$^\dag$). Moreover, as a result of the symmetry unification, classes AI and $\mathrm{D}^{\dag}$, and classes AII and $\mathrm{C}^{\dag}$, are equivalent to each other. Adding class A,  which has no symmetry at all, we have in total the 9 classes as listed in Table~\ref{table1} for non-Hermitian classes where up to one symmetry is relevant (the other 29 classes have more than one symmetry, such as PHS and TRS). We note that the entire 10 Altland-Zirnbauer symmetry classes for Hermitian matrices~\cite{Altland97} ramify into 38 symmetry classes for non-Hermitian matrices~\cite{Kawabata19S}. In the following, we describe basic properties of non-Hermitian matrices for each of the above-mentioned 9 classes.

%%%%%
\subsection{Class A}

Matrices in class A are not constrained by any symmetry and thus include most general non-Hermitian matrices. The probability distribution of a Gaussian ensemble is given as
\begin{align}
P \left( H \right) dH
&\propto e^{-\beta \mr{Tr}[H^{\dag} H]} dH\nonumber\\
&\propto \exp \left[ -\beta \sum_{i, j} |H_{ij}|^2 \right] \prod_{i, j}dH_{ij}dH_{ij}^*,
\end{align}
where $H_{ij}$ is the element in the $i$th row and the $j$th column of the matrix $H$.
On the other hand, for a Bernoulli ensemble, each matrix element is randomly chosen as
\begin{equation}
  H_{ij} = \left\{ \begin{array}{ll}
    1+i & \text{with probability 1/4}; \\
    1-i &  \text{with probability 1/4}; \\
    -1+i &  \text{with probability 1/4};\\
    -1-i &  \text{with probability 1/4}.
  \end{array} \right.
\end{equation}

%%%%%
\subsection{Class AI and class $\text{D}^{\dag}$}

Matrices in class AI respect TRS defined by
\begin{equation}
{\cal T}_{+} H^{*} {\cal T}_{+}^{-1} = H,\quad
{\cal T}_{+} {\cal T}_{+}^{*} = +1,
	\label{eq: app I - AI}
\end{equation}
where ${\cal T}_{+}$ is a unitary matrix (i.e., ${\cal T}_{+} {\cal T}_{+}^{\dag} = {\cal T}_{+}^{\dag} {\cal T}_{+} = 1$). In the presence of TRS, we have
\begin{equation}
H \left( {\cal T}_{+} \phi_{\alpha}^{*} \right)
= {\cal T}_{+} H^{*} \phi_{\alpha}^{*}
= E_{\alpha}^{*} \left( {\cal T}_{+} \phi_{\alpha}^{*} \right).
\end{equation}
Hence, ${\cal T}_{+} \phi_{\alpha}^{*}$ is also an eigenvector of $H$ with its eigenvalue $E_{\alpha}^{*}$, and eigenvalues form $( E_{\alpha}, E_{\alpha}^{*} )$ pairs in general.
Note that the eigenvalue remains real if the corresponding eigenvector satisfies ${\cal T}_{+} \phi_{\alpha}^{*} \propto \phi_{\alpha}$.
 Similarly, matrices in class $\text{D}^{\dag}$ respect the Hermitian conjugate of PHS, which we denote by $\text{PHS}^{\dag}$:
\begin{equation}
{\cal T}_{-} H^{*} {\cal T}_{-}^{-1} = - H,\quad
{\cal T}_{-} {\cal T}_{-}^{*} = +1,
	\label{eq: app I - Dd}
\end{equation}
which leads to $( E_{\alpha}, - E_{\alpha}^{*} )$ pairs in the complex plane. Notably, symmetry classes AI and $\text{D}^{\dag}$ are equivalent to each other~\cite{Kawabata18T}. In fact, when a non-Hermitian matrix $H$ satisfies Eq.~(\ref{eq: app I - AI}) and belongs to class AI, another non-Hermitian matrix $iH$ satisfies Eq.~(\ref{eq: app I - Dd}) and belongs to class $\text{D}^{\dag}$.
In particular, the level-spacing distributions are the same for classes AI and $\text{D}^{\dag}$, provided that the spectrum is rotated by 90 degrees in the complex plane.

For Gaussian ensembles, we can assume that $\mathcal{T}_+$ is the identity operator without loss of generality, which leads to
\begin{equation}\label{eq:realmat}
H_{ij} = H_{ij}^{*}.
\end{equation}
Thus, matrices in class AI can be represented  by real non-Hermitian matrices.
Indeed, the condition (\ref{eq: app I - AI}) means that $U^{-1}HU$ is real, where $U=\sqrt{\mc{T}_+}$ is unitary.
Since $P(H)dH=P(U^{-1}HU)d(U^{-1}HU)$ for Gaussian ensembles, we can consider the ensemble of real matrices in Eq.~(\ref{eq:realmat}).
Then we can consider the probability distribution of a Gaussian ensemble  given as
\begin{equation}
P \left( H \right) dH
\propto \exp \left[ -\beta \sum_{i, j} H_{ij}^2 \right] \prod_{i, j} dH_{ij}.
\end{equation}

While a Bernoulli ensemble depends on the explicit form of $\mc{T}_+$, we here consider $\mc{T}_+=1$.
Then each matrix element is randomly chosen as
\begin{equation}
  H_{ij} = \left\{ \begin{array}{ll}
    1 &  \text{with probability 1/2};\\
    -1 &  \text{with probability 1/2}.
  \end{array} \right.
\end{equation}
　

%%%%%
\subsection{Class AII and class $\text{C}^{\dag}$}

Matrices in class AII respect TRS defined by
\begin{equation}
{\cal T}_{+} H^{*} {\cal T}_{+}^{-1} = H,\quad
{\cal T}_{+} {\cal T}_{+}^{*} = -1.
	\label{eq: app I - AII}
\end{equation}
In analogy with class AI, eigenvalues form $( E_{\alpha}, E_{\alpha}^{*} )$ pairs in general. In addition, class AII is equivalent to class $\text{C}^{\dag}$, whose matrices respect $\text{PHS}^{\dag}$:
\begin{equation}
{\cal T}_{-} H^{*} {\cal T}_{-}^{-1} = - H,\quad
{\cal T}_{-} {\cal T}_{-}^{*} = -1.
	\label{eq: app I - Cd}
\end{equation}
In contrast to the Hermitian case and class $\text{AII}^{\dag}$ described below, only those eigenvalues that lie on the real axis can be two-fold degenerate through formation of Kramers pairs.
However, eigenvalues away from the real axis are not, in general, degenerate because they do not form Kramers pairs.
However, Kramers pairs on the real axis do not exist for almost all random matrices, since eigenvalues for generic matrices exhibit level repulsions.

We choose ${\cal T}_{+}$ as a matrix
\aln{
\Sigma^{y}=\sigma^y\otimes\mbb{I}_{{\frac{N}{2}}\times{\frac{N}{2}}},
}
 which again allows us to describe the probability distribution for Gaussian ensembles for general $\mc{T}_+$ with $\mc{T}_+\mc{T}_+^*=-1$. In this case, we obtain
\begin{equation}
H = \mbb{I}_{2\times 2} \otimes a + i \sigma^{x} \otimes b + i \sigma^{y} \otimes c + i \sigma^{z} \otimes d,
\end{equation}
where $a$, $b$, $c$, and $d$ are real $N/2\times N/2$ non-Hermitian matrices. The probability distribution of a Gaussian ensemble is then given as
\begin{align}
P \left( H \right) dH
\propto &\exp \left[ -\beta \sum_{i, j} \left( a_{ij}^2 + b_{ij}^2 + c_{ij}^2 + d_{ij}^2 \right) \right] \nonumber\\
&\quad\quad\quad\quad\quad\quad\quad\quad\times\prod_{i, j} da_{ij}db_{ij}dc_{ij}dd_{ij}.
\end{align}
On the other hand, for a Bernoulli ensemble, each element is randomly chosen as
\begin{equation}
  a_{ij},b_{ij},c_{ij},d_{ij} = \left\{ \begin{array}{ll}
    1 &  \text{with probability 1/2};\\
    -1 &  \text{with probability 1/2}.
  \end{array} \right.
\end{equation}

%%%%%
\subsection{Class $\text{AI}^{\dag}$}

Matrices in class $\text{AI}^{\dag}$ respect the Hermitian conjugate of TRS, i.e., $\text{TRS}^{\dag}$, defined by
\begin{equation}
{\cal C}_{+} H^{T} {\cal C}_{+}^{-1} = H,\quad
{\cal C}_{+} {\cal C}_{+}^{*} = +1,
	\label{eq: app I - AId}
\end{equation}
where ${\cal C}_{+}$ is a unitary matrix (i.e., ${\cal C}_{+} {\cal C}_{+}^{\dag} = {\cal C}_{+}^{\dag} {\cal C}_{+} = 1$). Noting that the transpose of the eigenequation~(\ref{eq: app I - eigenequation}) gives
\begin{equation}
H^{T} \chi_{\alpha}^{*}
= E_\alpha \chi_{\alpha}^{*},
\end{equation}
we have, in the presence of $\text{TRS}^{\dag}$,
\begin{equation}
H \left( {\cal C}_{+} \chi_{\alpha}^{*} \right)
= {\cal C}_{+} H^{T} \chi_{\alpha}^{*}
= E_{\alpha} \left( {\cal C}_{+} \chi_{\alpha}^{*} \right).
\end{equation}
Hence, ${\cal C}_{+} \chi_{\alpha}^{*}$ is also an eigenvector of $H$ having the same eigenvalue $E_{\alpha}$ as $\phi_{\alpha}$. Thus, if there is no degeneracy, the constraint
\begin{equation}
{\cal C}_{+} \chi_{\alpha}^{*} \propto \phi_{\alpha}
\end{equation}
is imposed on the right and left eigenvectors. Importantly, this symmetry constraint is imposed for all the eigenvectors in the entire complex plane in stark contrast to class AI, which leads to a new universality class of the level-spacing distribution as demonstrated in the main text.

We assume that $\mathcal{C}_+$ is the identity (which again allows us to describe the probability distribution for Gaussian ensembles for general $\mc{C}_+$ with $\mc{C}_+\mc{C}_+^*=1$), which leads to
\begin{equation}\label{tenti}
H_{ij} = H_{ji}.
\end{equation}
Thus, matrices in class $\text{AI}^{\dag}$ can be represented  by symmetric non-Hermitian matrices. The probability distribution of a Gaussian ensemble is then given as
\begin{align}
P \left( H \right) dH
\propto& \exp \left[ -\beta \left( \sum_i |H_{ii}|^2+ \sum_{i> j} 2|H_{ij}|^2 \right) \right]\nonumber\\
&\quad\quad\quad\quad\quad\quad\quad\quad\times\prod_{i\geq j} dH_{ij} dH_{ij}^*.
\end{align}
On the other hand, for a Bernoulli ensemble, each matrix element is randomly chosen as
\begin{equation}
  H_{ij} = \left\{ \begin{array}{ll}
    1+i & \text{with probability 1/4}; \\
    1-i &  \text{with probability 1/4}; \\
    -1+i &  \text{with probability 1/4};\\
    -1-i &  \text{with probability 1/4}
  \end{array} \right.
\end{equation}
under the constraint (\ref{tenti}).

%%%%%
\subsection{Class $\text{AII}^{\dag}$}

Matrices in class $\text{AII}^{\dag}$ respect $\text{TRS}^{\dag}$ defined by
\begin{equation}
{\cal C}_{+} H^{T} {\cal C}_{+}^{-1} = H,\quad
{\cal C}_{+} {\cal C}_{+}^{*} = -1.
	\label{eq: app I - AIId}
\end{equation}
Importantly, there is a non-Hermitian generalization of the Kramers degeneracy theorem in class $\text{AII}^{\dag}$~\cite{Esaki11, Sato12T}. In fact, from ${\cal C}_{+}^{T} {\cal C}_{+}^{-1} = -1$, we have
\begin{equation}
\chi_{\alpha}^{\dag} {\cal C}_{+} \chi_{\alpha}^{*}
= \left( \chi_{\alpha}^{\dag} {\cal C}_{+} \chi_{\alpha}^{*} \right)^{T}
= \chi_{\alpha}^{\dag} {\cal C}_{+}^{T} \chi_{\alpha}^{*}
= - \chi_{\alpha}^{\dag} {\cal C}_{+} \chi_{\alpha}^{*},
\end{equation}
which leads to $\chi_{\alpha}^{\dag} {\cal C}_{+} \chi_{\alpha}^{*} = 0$. Thus, the eigenvectors $\phi_{\alpha}$ and ${\cal C}_{+} \chi_{\alpha}^{*}$ are biorthogonal to each other and linearly independent of each other. This independence implies that all the eigenvectors are at least two-fold degenerate in the presence of $\text{TRS}^{\dag}$ with ${\cal C}_{+} {\cal C}_{+}^{*} = -1$. Since all the eigenvectors are subject to this non-Hermitian extension of the Kramers degeneracy, class $\text{AII}^{\dag}$ is  sharply contrasted with class AII in non-Hermitian systems.

We choose ${\cal C}_{+}$ as the matrix $\Sigma^{y}$, which  allows us to  describe the probability distribution for Gaussian ensembles for general $\mc{C}_+$ with $\mc{C}_+\mc{C}_+^*=-1$. We then have
\begin{equation}
H = \begin{pmatrix}
a & b \\ c & d
\end{pmatrix},
%I \otimes a + \sigma_{x} \otimes b + \sigma_{y} \otimes c + \sigma_{z} \otimes d,
	\label{eq: app I - AIId - rep}
\end{equation}
where non-Hermitian matrices $a$, $b$, $c$, and $d$ satisfy
\begin{equation}\label{AIIdag}
a = d^{T},\quad
b = - b^{T},\quad
c = - c^{T}.
\end{equation}
The probability distribution of a Gaussian ensemble is then given as
\begin{equation} \begin{split}
P& \left( H \right) dH\\
\propto& \exp \left\{ -2\beta \left[ \sum_{i\geq j} | a_{ij}|^2+ \sum_{i>j} \left( | b_{ij}|^2+| c_{ij}|^2+| d_{ij}|^2 \right) \right] \right\} \\
&\quad\quad
\times\prod_{i\geq j} da_{ij}da_{ij}^*
\prod_{i>j}db_{ij}db_{ij}^*dc_{ij}dc_{ij}^*dd_{ij}dd_{ij}^*.
\end{split} \end{equation}
On the other hand, for a Bernoulli ensemble, each element is randomly chosen as
\begin{equation}
  a_{ij},b_{ij},c_{ij},d_{ij} = \left\{ \begin{array}{ll}
    1+i & \text{with probability 1/4}; \\
    1-i &  \text{with probability 1/4}; \\
    -1+i &  \text{with probability 1/4};\\
    -1-i &  \text{with probability 1/4}
  \end{array} \right.
\end{equation}
under the constraint (\ref{AIIdag}).

%%%%%
\subsection{Class D}

Matrices in class D respect PHS defined by
\begin{equation}
{\cal C}_{-} H^{T} {\cal C}_{-}^{-1} = - H,\quad
{\cal C}_{-} {\cal C}_{-}^{*} = +1,
	\label{eq: app I - D}
\end{equation}
where ${\cal C}_{-}$ is a unitary matrix (i.e., ${\cal C}_{-} {\cal C}_{-}^{\dag} = {\cal C}_{-}^{\dag} {\cal C}_{-} = 1$). In analogy with classes $\text{AI}^{\dag}$ and $\text{AII}^{\dag}$, we have
\begin{equation}
H \left( {\cal C}_{-} \chi_{\alpha}^{*} \right)
= - {\cal C}_{-} H^{T} \chi_{\alpha}^{*}
= - E_{\alpha} \left( {\cal C}_{-} \chi_{\alpha}^{*} \right).
\end{equation}
Hence, ${\cal C}_{-} \chi_{\alpha}^{*}$ is also an eigenvector of $H$ with its eigenvalue $-E_{\alpha}$, and eigenvalues form $\left( E_{\alpha}, - E_{\alpha} \right)$ pairs in general. Thus, PHS only makes pairs of eigenvalues $\left( E_{\alpha}, - E_{\alpha} \right)$ and imposes no local constraints on generic eigenvectors away from the zero eigenvalue, which is similar to TRS and $\text{PHS}^{\dag}$ and in contrast with $\text{TRS}^{\dag}$.

We assume that $\mathcal{C}_{-}$ is the identity operator, which  allows us to describe the probability distribution for Gaussian ensembles for general $\mc{C}_-$ with $\mc{C}_-\mc{C}_-^*=1$. We then have
\begin{equation}\label{atenti}
H_{ij} = - H_{ji}.
\end{equation}
Thus, matrices in class D can be represented  by antisymmetric non-Hermitian matrices. The probability distribution of a Gaussian ensemble is then given as
\begin{equation}
P \left( H \right) dH
\propto \exp \left[ -2\beta\sum_{i> j} |H_{ij}|^2 \right]
\prod_{i>j} dH_{ij} dH_{ij}^*.
\end{equation}
On the other hand, for a Bernoulli ensemble, each matrix element is randomly chosen as
\begin{equation}
  H_{ij} = \left\{ \begin{array}{ll}
    1+i & \text{with probability 1/4}; \\
    1-i &  \text{with probability 1/4}; \\
    -1+i &  \text{with probability 1/4};\\
    -1-i &  \text{with probability 1/4}
  \end{array} \right.
\end{equation}
subject to the constraint~(\ref{atenti}).

%%%%%
\subsection{Class C}

Matrices in class C respect PHS defined by
\begin{equation}
{\cal C}_{-} H^{T} {\cal C}_{-}^{-1} = - H,\quad
{\cal C}_{-} {\cal C}_{-}^{*} = -1.
	\label{eq: app I - C}
\end{equation}
In analogy with class D, eigenvalues form $\left( E_{\alpha}, - E_{\alpha} \right)$ pairs in general.

We choose ${\cal C}_{-}$ as the  matrix $\Sigma^{y}$, which allows us to describe the probability distribution for Gaussian ensembles for general $\mc{C}_-$ with $\mc{C}_-\mc{C}_-^*=-1$. We then have Eq.~(\ref{eq: app I - AIId - rep}) with
\begin{equation}\label{C}
a = - d^{T},\quad
b = b^{T},\quad
c = c^{T}.
\end{equation}
The probability distribution of a Gaussian ensemble is then given as
\begin{equation} \begin{split}
P &\left( H \right) dH\\
\propto& \exp \lrm{ -2\beta  \sum_{i\geq j} \left( | a_{ii} |^2+\frac{| b_{ii} |^2}{2}+\frac{| c_{ii} |^2}{2} \right) }\\
&\times\exp\lrm{-2\beta \sum_{i>j} \left( | a_{ij} |^2+| b_{ij} |^2+| c_{ij} |^2+|d_{ij} |^2 \right) } \\
&
\times\prod_{i\geq j} da_{ij} da_{ij}^{*} db_{ij} db_{ij}^{*}
\prod_{i>j}  dc_{ij} dc_{ij}^{*} dd_{ij} dd_{ij}^{*}.
\end{split} \end{equation}
On the other hand, for a Bernoulli ensemble, each element is randomly chosen as
\begin{equation}
  a_{ij},b_{ij},c_{ij},d_{ij} = \left\{ \begin{array}{ll}
    1+i & \text{with probability 1/4}; \\
    1-i &  \text{with probability 1/4}; \\
    -1+i &  \text{with probability 1/4};\\
    -1-i &  \text{with probability 1/4}
  \end{array} \right.
\end{equation}
subject to the constraint~(\ref{C}).

%%%%%
\subsection{Class AIII}

Matrices in class AIII respect CS defined by
\begin{equation}
\Gamma H^{\dag} \Gamma^{-1} = - H,\quad
\Gamma^{2} = 1,
	\label{eq: app I - AIII}
\end{equation}
where $\Gamma$ is a unitary matrix (i.e., $\Gamma \Gamma^{\dag} = \Gamma^{\dag} \Gamma = 1$). In the presence of CS, we have
\begin{equation}
H \left( \Gamma \chi_{\alpha} \right)
= - \Gamma H^{\dag} \chi_{\alpha}
= - E_{\alpha}^{*} \left( \Gamma \chi_{\alpha} \right).
\end{equation}
Hence, $\Gamma \chi_{\alpha}$ is also an eigenvector of $H$ with its eigenvalue $- E_{\alpha}^{*}$, and eigenvalues form $( E_{\alpha}, - E_{\alpha}^{*} )$ pairs in general. In analogy with the equivalence between TRS and $\text{PHS}^{\dag}$~\cite{Kawabata18T}, CS is equivalent to pseudo-Hermiticity~\cite{Mostafazadeh02} which is defined by the presence of the unitary matrix $\eta$ such that
\begin{equation}
\eta H^{\dag} \eta^{-1} = H,\quad
\eta^{2} = 1.
	\label{eq: app I - pH}
\end{equation}
This condition implies the presence of $( E_{\alpha}, E_{\alpha}^{*} )$ pairs in the complex plane. In fact, when a non-Hermitian matrix $H$ satisfies Eq.~(\ref{eq: app I - AIII}) and respects CS, another non-Hermitian matrix $iH$ satisfies Eq.~(\ref{eq: app I - pH}) and respects pseudo-Hermiticity.
We note that unlike Hermitian systems, CS is  distinct from sublattice symmetry (SLS) defined by Eq.~(\ref{eq: app I - SLS}).

We choose $\Gamma$ to be a  matrix
\aln{
\Sigma^z=\sigma\otimes\mbb{I}_{\frac{N}{2}\times\frac{N}{2}}
}
 to obtain a  nontrivial result, which leads to Eq.~(\ref{eq: app I - AIId - rep}) with
\begin{equation}\label{AIII2}
a = - a^{\dag},\quad
b = c^{\dag},\quad
c = b^{\dag},\quad
d = - d^{\dag}.
\end{equation}
We note that $H$ always reduces to an anti-Hermitian matrix when we take a special choice $\Gamma = \mbb{I}$, which trivially reduces to class A in Hermitian systems and is not considered here.
The probability distribution of a Gaussian ensemble is then given as
\begin{equation} \begin{split}
P &\left( H \right) dH\\
\propto& \exp \lrm{ - \beta \sum_{i} \left( | a_{ii} |^2 + | d_{ii} |^2 + 2\,|b_{ii}|^2 \right) }\\
&\times\exp\lrm{ -2\beta\sum_{i>j} \left( |a_{ij}|^2+|b_{ij}|^2+|c_{ij}|^2+|d_{ij}|^2 \right)}\\
&
\times\prod_{i\geq j} da_{ij} dd_{ij} db_{ij}db_{ij}^*\prod_{i>j}dc_{ij}dc_{ij}^*dd_{ij}dd_{ij}^*.
\end{split} \end{equation}
On the other hand, for a Bernoulli ensemble, each element is randomly chosen as
\begin{equation}\label{AIII}
  a_{ij},b_{ij},c_{ij},d_{ij} = \left\{ \begin{array}{ll}
    1+i & \text{with probability 1/4}; \\
    1-i &  \text{with probability 1/4}; \\
    -1+i &  \text{with probability 1/4};\\
    -1-i &  \text{with probability 1/4}
  \end{array} \right.
\end{equation}
subject to the constraint~(\ref{AIII2}).

%%%%%
\subsection{Class $\text{AIII}^{\dag}$}

Matrices in class $\text{AIII}^{\dag}$ respect SLS defined by
\begin{equation}
{\cal S} H {\cal S}^{-1} = - H,\quad
{\cal S}^{2} = 1,
	\label{eq: app I - SLS}
\end{equation}
where ${\cal S}$ is a unitary matrix (i.e., ${\cal S} {\cal S}^{\dag} = {\cal S}^{\dag} {\cal S} = 1$). In the presence of SLS, we have
\begin{equation}
H \left( {\cal S} \phi_{\alpha} \right)
= - {\cal S} H \phi_{\alpha}
= - E_{\alpha} \left( {\cal S} \phi_{\alpha} \right).
\end{equation}
Hence, ${\cal S} \phi_{\alpha}$ is also an eigenvector of $H$ with its eigenvalue $- E_{\alpha}$, and eigenvalues form $( E_{\alpha}, - E_{\alpha} )$ pairs in general.

We choose ${\cal S}$ as the  matrix $\Sigma^{z}$ to obtain a nontrivial result, which leads to Eq.~(\ref{eq: app I - AIId - rep}) with
\begin{equation}
a = d = 0.
\end{equation}
In analogy with class AIII, we do not consider the special choice ${\cal S} = \mbb{I}$, which leads to the trivial case $H = 0$.
The probability distribution of a Gaussian ensemble is then given as
\begin{equation}
P \left( H \right) dH
\propto \exp \left[ -\beta \sum_{i, j} \left( | b_{ij} |^2 + | c_{ij} |^2 \right) \right] \prod_{i, j} db_{ij} dc_{ij}.
\end{equation}
On the other hand, for a Bernoulli ensemble, each element is randomly chosen as
\begin{equation}
  b_{ij},c_{ij}= \left\{ \begin{array}{ll}
    1+i & \text{with probability 1/4}; \\
    1-i &  \text{with probability 1/4}; \\
    -1+i &  \text{with probability 1/4};\\
    -1-i &  \text{with probability 1/4}.
  \end{array} \right.
\end{equation}

%\section{Bernoulli distributions}
%We introduce

%\section{Degenerate perturbation theory and level correlations}

\section{Proof of Eq.~\eqref{important} and its meaning}~\label{appB}
In this appendix, we compute the level-spacing probability distributions of $H_\mr{small, A}$, $H_\mr{small, AI^\dag}$, and $H_\mr{small, AII^\dag}$.
 First, we express the level spacing $s$ in terms of the following stochastic variable:
 \begin{equation}
  X_f = z_1^2 + \dotsb + z_f^2,
   \label{192758_17Dec18}
 \end{equation}
 where $z_1,\cdots,z_f$ are independently and identically distributed (i.i.d.) complex variables under the Gaussian distribution $P(z)\propto e^{-|{z}|^2}$.

We show below the following relation
 \begin{equation}
  s \propto \lrv{X_f}^{1/2},
   \label{192505_11Dec18}
 \end{equation}
 where $f=2, 3$, and $5$ for classes  AI$^\dag$, A, and AII$^\dag$, respectively.
 After showing (\ref{192505_11Dec18}), we derive the probability distribution function of $\lrv{X_f}^2$, from which the level-spacing distribution follows.

 \subsection{Derivation of Eq.~(\ref{192505_11Dec18})}
 For classes A, AI$^\dag$, and AII$^\dag$, the set of all the matrices with the corresponding symmetry forms a complex vector space that is closed under  Hermitian conjugation.
 Therefore, a random $n\times n$ matrix $H$ can take the following form:
 \begin{equation}
  H = z_0 I + \sum_{j=1}^{f} z_j L_j, \qquad
   z_0, z_1,\dotsc, z_f \in \mathbb{C},
   \label{190026_11Dec18}
 \end{equation}
 where $L_1,\dotsc, L_f$ are Hermitian matrices satisfying
 \begin{equation}
  \Tr L_j = 0, \qquad
   \Tr L_iL_j = n \delta_{ij}.
   \label{185759_11Dec18}
 \end{equation}
% The matrix $H_\mathrm{small}$ can be expressed with the linear basis $\{I, L_1,\dotsc, L_f\}$:
% \begin{equation}
%  H_\mathrm{small} = z_0 I + \sum_{j=1}^{f} z_j L_j, \qquad
%   z_0, z_1,\dotsc, z_f \in \mathbb{C}.
%   \label{190026_11Dec18}
% \end{equation}
 We regard the coefficients $z_0,\dotsc,z_f$ as random variables rather than matrix entries.  The condition \eqref{185759_11Dec18} implies that $z_0,z_1,\ldots,z_f$ are i.i.d.\@ Gaussian variables:
 \begin{equation}
  P(H) \propto e^{-\beta \mr{Tr}[H^\dagger H]}
   = \exp [-n\beta(\lrv{z_0}^2+\lrv{z_1}^2+\dotsb+\lrv{z_f}^2)].
   \label{193612_17Dec18}
 \end{equation}
 Since the term $z_0 \mbb{I}$ does not affect the level spacing, we eliminate $z_0$ from Eq.~\eqref{190026_11Dec18} and instead consider the traceless random matrix:
 \begin{equation}
  \tilde{H} = \sum_{j=1}^{f} z_j L_j.
   \label{191020_11Dec18}
 \end{equation}

 Now we consider $H_\mr{small, A}$, $H_\mr{small, AI^\dag}$, and $H_\mr{small, AII^\dag}$ defined in Eq.~\eqref{rmattwo}.
 In each of these matrices, the matrix basis $\{L_1, \ldots, L_f\}$ can be taken as listed in Table~\ref{191459_11Dec18}.
 The matrices in the basis become spinor matrices characterized by the anticommutation relation
 $\frac{1}{2}(L_iL_j+L_jL_i) = \delta_{ij}I$.
 \begin{table}[tb]
  \centering
  \caption{\label{191459_11Dec18}%
  Examples of matrix bases for classes  AI$^\dag$, A, and AII$^\dag$.
  Note that $\sigma^{x,y,z}$ are Pauli matrices and that $\gamma^{1,2,3,4,5}$ are the Dirac matrices according to the notation in Ref.~\cite{Sakurai67}.
  }
  \smallskip
  \begin{tabular}{ccc}
   \hline   \hline
   ~~Class~~    & ~~$f$~~ & ~~Basis~~ \\
   \hline
   AI$^\dag$        &    2   & $\sigma^x, \sigma^z$ \\
   A  &    3   & $\sigma^x, \sigma^y, \sigma^z$ \\
   AII$^\dag$ &    5   & ~~$\gamma^1, \gamma^2, \gamma^3, \gamma^4, \gamma^5$~~ \\
   \hline   \hline
  \end{tabular}
 \end{table}
 As a consequence, the traceless matrix $\tilde{H}_\mathrm{small}$ obtained from $H_\mr{small}$ satisfies
 \begin{equation}
  \tilde{H}_\mathrm{small}^2 = (z_1^2+\dotsb+z_f^2)\mbb{I}.
   \label{172454_17Dec18}
 \end{equation}

 At the same time, $\tilde{H}_\mathrm{small}$ has only two distinct eigenvalues $(\lambda,-\lambda)$ of the same multiplicity.
 Therefore, we also have the following equality:
 \begin{equation}
  \tilde{H}_\mathrm{small}^2 = \lambda^2\mbb{I}.
   \label{172459_17Dec18}
 \end{equation}
 Comparing Eqs.~\eqref{172454_17Dec18} and \eqref{172459_17Dec18}, the level spacing of $\tilde{H}_\mathrm{small}$ (and hence of $H_\mathrm{small}$) is found to be described by the following quantity:
 \begin{equation}
  s = 2\lrv{\lambda} = 2\sqrt{n\beta}\lrv{X_f}^{1/2},
 \end{equation}
 where the factor $\sqrt{n\beta}$ arises from the difference in the scaling between the random variables in \eqref{193612_17Dec18} and those in \eqref{192758_17Dec18}.

% A similar discussion for the Hermitian cases shows that the sum of the squares of $f$ real variables leads to a factor given by the chi-squared distribution $\chi_f^2(X_f')$ for $X_f'=s^2\in\mbb{R}$.
% In this case, the geometrical factor is $s^1$ by transforming $X_f'$ into $s$, instead of $s^3$ in the non-Hermitian cases.
% Then we obtain $p_{\mr{WD}}(s)\propto s\chi_f^2(s^2)$, where the degree $f$ of the chi-squared distribution $\chi_f^2(s^2)= \mr{O}(s^{f-2})$ changes the degree of level repulsion.
% To our best knowledge, such a unified interpretation of level-spacing distributions for Hermitian matrices from the chi-squared distribution has never been presented before.
% Our results indicate that the degree of freedom, which changes depending on the symmetry, not only determines level repulsion but also other aspects of $p_\mr{small}$, such as the position of the peak or the length of the tail, solely through $\chi_f^2(X_f')$.
% Since $p(s)$ for large matrices is well approximated by $p_\mr{small}$ in the  Hermitian case, it is also approximately described by $\chi_f^2(X_f')$.
 A similar argument applies to the analysis of Hermitian random matrices if we use real Gaussian variables instead of complex ones.
 In this case, the stochastic variable $X_f$ is replaced by $X_f' = x_1^2 + \dotsb + x_f^2$ with real Gaussian variables $x_1,\ldots,x_f$.
 Then, $X_f'$ is subject to the chi-squared distribution: $P(X_f' = r) = \chi_f^2(r) \propto r^{f/2-1}e^{-r}$.
 Therefore, the level-spacing distribution is the transformed chi-squared distribution,
 \aln{
 p_\mr{small}(s) = 2s\chi_f^2(s^2) \propto s^{f-1}e^{-(Cs)^2},
 }
  where the level repulsion $O(s^{f-1})$ depends on the degree $f$ of the chi-squared distribution.
 Since $p(s)$ for large matrices is well approximated by $p_\mr{small}$ in the Hermitian case, it is also approximately described by $\chi_f^2(X_f')$.

 In the following, the distribution of $X_f$ for non-Hermitian matrices is shown to be the $K$-distribution.  The degree of the level repulsions is confirmed to be independent of $f \ge 2$ except for a logarithmic correction at $f=2$.

 \subsection{Distribution function of {$\lrv{X_f}^2$}}
 % Since the variance of the Gaussian variables $z_j$ only affects the scaling of the level spacing, we may set it to unity with losing the generality.

 The probability density of the squared modulus $\lrv{X_f}^2$ is given by
 \begin{equation}
  P(\lrv{X_f}^2 = \rho) =
   \int_{}\:\delta\left(\rho - \lrv{z_1^2 + \cdots + z_f^2}^2\right)\pi^{-f}e^{-||{\mathbf{z}}||^2} d^{2f}\mathbf{z},
   \label{132423_12Dec18}
 \end{equation}
 where $\mathbf{z} = (z_1,\ldots,z_f)$ is a complex vector.
 Let  $\mathbf{u}$ and $\mathbf{v}$ be the real and imaginary parts of $\mathbf{z}$ ($\mathbf{z}=\mathbf{u} + i\mathbf{v}$), respectively.  Then,  $X_f$ is expressed in terms of the magnitudes $u, v$ of $\mathbf{u}, \mathbf{v}$ and the angle $\theta$ between them as
 \begin{align}
  X_f &= ||{\mathbf{u}}||^2 - ||{\mathbf{v}}||^2 + 2i\mathbf{u}\cdot \mathbf{v}
  = u^2 - v^2 + 2iuv\cos \theta.
  \end{align}
 Thus,
 \begin{align}
 \lrv{X_f}^2 &= u^4 + v^4 + 2u^2v^2 (2\cos^2 \theta - 1).
 \end{align}
Substituting $(U, V, q) = (u^2, v^2,\cos \theta)$ for $(u,v,\theta)$, we obtain
 \begin{equation}
  \lrv{X_f}^2 = U^2 + V^2 + 2UV (2q^2 - 1).
 \end{equation}
Associated with this change of variables, the measure $d^{2f}\mathbf{z}$ is transformed as
 \begin{align}
  d^{2f}\mathbf{z} &= d^f\mathbf{u} d^f\mathbf{v}\notag\\ &=
  \frac{2\pi^{f/2}}{\Gamma(f/2)}\frac{2\pi^{(f-1)/2}}{\Gamma\bigl((f-1)/2\bigr)}
  u^{f-1}du v^{f-1}dv \sin^{f-2} \theta d\theta \notag\\
  &=  \frac{2^{f}\pi^{f-1}}{\Gamma(f-1)}
  u^{f-1}du v^{f-1}dv \sin^{f-2} \theta d\theta \notag\\
  &=  \frac{2^{f-2}\pi^{f-1}}{\Gamma(f-1)}(UV)^{(f-2)/2}dUdV(1-q^2)^{(f-3)/2}dq,
 \end{align}
 where the duplication formula $\Gamma(x)\Gamma\bigl(x+1/2) = 2^{1-2x}\pi^{1/2}\Gamma(2x)$ is used.  Therefore, Eq.~\eqref{132423_12Dec18} becomes
\begin{widetext}
 \begin{align}
  P(\lrv{X_f}^2 = \rho) =\frac{1}{\pi \Gamma(f-1)}
  \int_0^\infty  \int_0^\infty  dUdV (4UV)^{\frac{f-2}{2}} e^{-(U+V)} \int_{-1}^1 (1-q^2)^{\frac{f-3}{2}}dq \delta\bigl(\rho-[U^2+V^2+2UV(2q^2 - 1)]\bigr).
 \end{align}
 \end{widetext}
 We integrate the delta function with respect to $q$.  For this purpose, we define
 \begin{equation}
  q_\rho = \biggl(\frac{\rho - (U-V)^2}{4UV}\biggr)^{1/2},
 \end{equation}
 which satisfies $0\le q_\rho\le 1$ if and only if $\lrv{U-V}\le \sqrt{\rho}\le U+V$.
 Then we obtain
 \begin{align}
  \delta&\bigl(\rho-[U^2+V^2+2UV(2q^2 - 1)]\bigr)\nonumber\\
  &= \frac{1}{8UV q_\rho}[\delta(q-q_\rho) + \delta(q+q_\rho)]
 \end{align}
 and
 \begin{align}
  &P(\lrv{X_f}^2 = \rho) \nonumber\\&= \frac{1}{\pi \Gamma(f-1)}
  \int \!\! dUdV
  (4UV)^{(f-2)/2} e^{-(U+V)}\frac{(1-q_\rho^2)^{(f-3)/2}}{4UVq_\rho}\nonumber \\
  &= \frac{1}{\pi\Gamma(f-1)} \int\!\! dUdV
  e^{-(U+V)}\frac{[(U+V)^2 - \rho]^{(f-3)/2}}{[\rho - (U-V)^2]^{1/2}},
 \end{align}
 where the integral is performed within the region satisfying $\lrv{U-V}\le \sqrt{\rho}\le U+V $ and we have used
 \begin{equation}
  \frac{(1-q_\rho^2)^{\frac{f-3}{2}}}{q_\rho} =
   \biggl(\frac{4UV}{\rho - (U-V)^2}\biggr)^{\frac{1}{2}}
   \biggl(\frac{(U+V)^2 - \rho}{4UV}\biggr)^{\frac{f-3}{2}}.
 \end{equation}
 Finally, we change the variables of integration from $(U,V)$ to $(x,y) = (U+V, U-V)$, obtaining
 \begin{align}
  P&(\lrv{X_f}^2 = \rho) \nonumber\\& =\frac{1}{2\pi\Gamma(f-1)}
  \int_{\sqrt{\rho}}^{\infty} dx \int_{-\sqrt{\rho}}^{\sqrt{\rho}} dy
  e^{-x}\frac{(x^2 - \rho)^{(f-3)/2}}{(\rho - y^2)^{1/2}}\nonumber \\
  &= \frac{1}{2\Gamma(f-1)}\int_{\sqrt{\rho}}^{\infty} dx
  e^{-x}(x^2 - \rho)^{(f-3)/2}.
  \label{135221_26Nov18}
 \end{align}
 Here we use the formula:
 \begin{align}
  \int_{r}^{\infty} dx e^{-x}(x^2 - r^2)^{\alpha-1/2}
   &= r^{2\alpha} \int_{0}^{\infty} dt e^{-r\cosh t}(\sinh t)^{2\alpha}\nonumber\\
   &= \frac{\Gamma(2\alpha)}{2^{\alpha-1}\Gamma(\alpha)} r^{\alpha} K_\alpha(r),
   \label{172337_13Dec18}
 \end{align}
 the proof of which will be provided at the end of this appendix.
 Substituting $\alpha = f/2-1$ and $r=\sqrt{\rho}$ in \eqref{172337_13Dec18}, we obtain the distribution function of $\lrv{X_f}^2$:
 \begin{align}
  P(\lrv{X_f}^2 = \rho)
  &= \frac{2^{-f/2}}{\Gamma(f/2)}\rho^{f/4-1/2}K_{f/2-1}(\sqrt{\rho}).
  \label{195117_17Dec18}
 \end{align}
 This probability distribution coincides with the $K$-distribution with shape parameters
 $(f/2,1)$~\cite{Jakeman78,Redding99}.
 In the limit of $\rho\to+0$, the right-hand side of \eqref{195117_17Dec18} converges to $\frac{1}{f-2}$ for $f>2$ and diverges as $O\bigl(\log (1/\rho)\bigr)$ at $f=2$.

 \subsection{Level-spacing distribution of $H_\mr{small}$}
 Since the level spacing $s$ is equal to $\rho^{1/4}$ times a scaling factor, the probability distribution of $s$ reads
 \begin{equation}
  p(s) \cong 4s^3 P(\lrv{X_f}^2=\rho)|_{\rho=s^4}
   = \frac{2^{2-f/2}}{\Gamma(f/2)}s^{f+1}K_{f/2-1}(s^2),
 \end{equation}
 where $\cong$ denotes the equivalence of stochastic variables that differ only by a constant factor.
 The level repulsion is $O(s^3)$ for $f>2$ and $O\bigl(s^3\log (1/s)\bigr)$ at $f=2$.

The distribution function is rescaled to the unit average spacing (i.e., $\int_0^\infty ds s p(s)=1$) by multiplying the scaling factor
\begin{align}
  C_f &= \int_{0}^{\infty} ds\frac{2^{2-f/2}}{\Gamma(f/2)}s^{f+2}K_{f/2-1}(s^2) \notag \\
 &= \int_0^{\infty}\frac{2^{-f/2}}{\Gamma(f/2)}\rho^{f/4-1/4}K_{f/2-1}(\sqrt{\rho}) \notag\\
 &= \frac{\Gamma(1/4)\Gamma(f/2+1/4)}{2\sqrt{2}\Gamma(f/2)},
 \end{align}
 where the last integration follows from the normalization of the $K$-distribution with shape parameters $\bigl(f/2+1/4,5/4\bigr)$~\cite{Redding99}.

 Thus, we have derived the level-spacing distribution for general $f$:
 \begin{align}
  p(s) &= \frac{2^{2-f/2}C_f^{f+2}}{\Gamma(f/2)}s^{f+1}K_{f/2-1}(C_f^2s^2) \notag \\
   &= \frac{1}{\mc{N}_f}(C_f s)^{f+1}K_{f/2-1}\bigl((C_fs)^2\bigr),
   \label{143807_2May19}
 \end{align}
 where we have set $\mc{N}_f = 2^{f/2-2}\Gamma\bigl(\frac{f}{2}\bigr) C_f^{-1}$,
 which leads to Eq.~\eqref{important}.
% \subsection{Different expressions of nearest-level-spacing distributions in Eq.~(2) in the main text}
% In Eq.~(2) in the main text, we express the nearest-level-spacing distributions in a unified manner as
% \aln{
%  p(s)=\frac{1}{\mc{N}_f}(C_fs)^{f+1}K_{\frac{f-2}{2}}((C_fs)^2),
% }
% where $f$ is the degree of freedom ($f=2$ for class AI$^\dag$, 3 for class A, and 5 for class AII$^\dag$),
% \aln{
% C_f=\frac{\Gamma\lrs{\frac{1}{4}}\Gamma\lrs{\frac{f}{2}+\frac{1}{4}}}{2\sqrt{2}\Gamma\lrs{\frac{f}{2}}}}
% and
% \aln{
% \mc{N}_f=2^{f/2-2}\Gamma\lrs{\frac{f}{2}}C_f^{-1}.

% }

 \subsection{Proof of Eq.~\eqref{172337_13Dec18}}
 \label{sec:addendum}
 Let us define
 \begin{align}
  f_1(r) &:= \int_{r}^{\infty} dx e^{-x}(x^2 - r^2)^{\alpha-1/2}, \\
  f_2(r) &:= r^{-\alpha}f_1(r), \\
  f_3(r) &:= r^{-2\alpha}f_1(r) = \int_{0}^{\infty} dt e^{-r\cosh t}(\sinh t)^{2\alpha}.
 \end{align}
 First, we derive a differential equation for $f_3(r)$:
 \begin{align}~\label{r3}
  r^2f_3(r) -& r^2f_3''(r) \notag\\
  &= r^2\int_0^{\infty} dte^{-r\cosh t}(1 - \cosh^2 t)(\sinh t)^{2\alpha} \notag \\
  &= r^2\int_0^{\infty} dte^{-r\cosh t}(\sinh t)^{2\alpha+2} \notag \\
  &= -r\int_0^{\infty} dt\biggl(\frac{d}{dr}e^{-r\cosh t}\biggr)(\sinh t)^{2\alpha+1} \notag \\
  &= r\int_0^{\infty} dte^{-r\cosh t}\biggl(\frac{d}{dr}(\sinh t)^{2\alpha+1}\biggr) \notag \\
  &= (2\alpha+1)r\int_0^{\infty} dte^{-r\cosh t}(\cosh t)(\sinh t)^{2\alpha} \notag \\
  &= (2\alpha+1)rf'_3(r).
 \end{align}
Substituting $f_3(r) = r^{-\alpha}f_2(r)$ in Eq.~(\ref{r3}), we obtain a differential equation for $f_2(r)$:
 \begin{align}
%  r^{-\alpha}[r^2f_2(r)-r^2f_2''(r)+2\alpha rf_2'(r)-\alpha(\alpha+1)f_2'(r)]
%  &= (2\alpha-1)r^{-\alpha}[rf_2(r) - \alpha f_2(r)] \\ \iff
  (r^2 + \alpha^2)f_2(r) + rf_2'(r) + r^2f_2''(r) &= 0,
 \end{align}
 which coincides with that of the modified Bessel equation.
 Here $f_1(r)$ vanishes at $r\to\infty$, so does $f_2(r)$, and therefore we can write $f_2(r) = CK_\alpha(r)$ with a constant $C$.
 This constant 
 can be determined by evaluating $f_1(0)$ in two ways:
 \begin{align}
  f_1(0) &= \int_0^{\infty} dxe^{-x}x^{2\alpha-1} = \Gamma(2\alpha), \\
  f_1(0) &= \lim_{r\to +0} Cr^\alpha K_\alpha(r) = 2^{\alpha-1}\Gamma(\alpha)C.
 \end{align}
Hence we obtain
\aln{
C = \frac{\Gamma(2\alpha)}{2^{\alpha-1}\Gamma(\alpha)}.
}

\bibliographystyle{apsrev4-1}
\bibliography{../../../../refer_them}

\end{document}